\documentclass[twocolumn,amssymb, nobibnotes, aps, prd, floatfix]{revtex4-2}

\usepackage{subfiles}
\usepackage{siunitx} 
\usepackage{graphicx}
\usepackage{dcolumn}
\usepackage{bm}
\usepackage{color,soul}
\usepackage{docs}%
\usepackage{float}  
\usepackage{lineno,hyperref}
\usepackage{atbegshi} 
\usepackage{capt-of}

\begin{document}

\title{Modeling of the positron sources: an experiment-based benchmarking}

\author{Fahad Alharthi}
\email{falharthi@kacst.gov.sa}
\thanks{\\
Also at: King Abdulaziz City for Science and Technology, KACST, Riyadh, Saudi Arabia.} 
\author{Iryna Chaikovska}
\author{Robert Chehab}
\author{Viktor Mytrochenko}
\affiliation{Universit\'e Paris-Saclay, CNRS/IN2P3, IJCLab,  91405 Orsay, France}

\author{ 
Fusashi Miyahara, 
Takuya Kamitani, 
Yoshinori Enomoto}%
\affiliation{High Energy Accelerator Research Organization (KEK), Oho 1-1, Tsukuba, Ibaraki, 305-0801, Japan}
\date{\today}

\begin{abstract}

High-intensity positron sources are critical for next-generation electron-positron colliders, where positron beam quality and characteristics directly impact the luminosity. Accurate modeling and validated simulation tools for positron sources are essential to optimize their performance. However, modeling a positron source tends to be complex, as it involves multiple interdependent stages, from positron production and capture dynamics to beam transport through the injector linac to the collider ring. A reliable simulation framework should integrate these processes to ensure efficient positron production and transport. In this work, we present a start-to-end simulation tool developed for positron sources modeling. The model was benchmarked against existing simulation tools and validated through experimental measurements conducted at the SuperKEKB positron source. Key operational parameters were systematically scanned to evaluate the simulation model performance, including the primary electron impact position on the target, solenoid field strength around the capture linac, and RF phase settings. The primary Figure-of-Merit for all validation tests was the positron yield at the end of the SuperKEKB positron capture section.
The simulation results demonstrate a very good agreement with experimental data and other simulation tools, confirming the model's reliability and establishing a framework for future positron source studies.

\end{abstract}

\maketitle

\section{Introduction}

With the current interest of the High-Energy Physics (HEP) community in precision studies of the Standard Model (SM) and the search for a possible extension beyond the SM (BSM), the need for high-luminosity lepton colliders is indispensable. At present, four possible projects are under comprehensive study: the Compact Linear Collider (CLIC) \cite{Aicheler2012}, or the Future Circular Collider (FCC-ee) \cite{Benedikt2019} at CERN, the International Linear Collider (ILC)\cite{Adolphsen:2013jya} in Japan, and the Circular Electron Positron Collider (CEPC) \cite{CEPCStudyGroup2023} in China. All these linear or circular colliders share the demand for high luminosity, typically on the order of $\sim 10^{35} \,\text{cm}^{-2} \text{s}^{-1}$. Achieving such high luminosities demands intense and low-emittance beams of electrons ($e^{-}$) and positrons ($e^{+}$). While high-intensity, low-emittance $e^{-}$ sources are well established, developing $e^{+}$ sources with similar characteristics remains a significant challenge for next-generation colliders. 
 
All $e^{+}$ sources in high-energy lepton colliders to date rely on pair production through the interaction of a high-intensity $e^{-}$ beam with a thick target (several radiation lengths, $X_0$) made of high-Z materials (e.g., tungsten, tantalum). Despite the high rate of $e^{+}$ production at the target, the challenge lies in capturing and transporting these $e^{+}$ through the injector while maintaining high transmission to the Damping Ring (DR). The performance of the $e^{+}$ source is typically measured by a key parameter known as the accepted yield, defined as the ratio of the number of $e^{+}$ accepted by the DR to the number of primary $e^{-}$ impinging on the target:  
\begin{equation}
    \eta = \frac{N_{e^{+}}}{N_{e^{-}}}.
\end{equation}

Maximizing the accepted yield requires a comprehensive start-to-end simulation framework that can optimize $e^{+}$ production, capture, and transport efficiency. As part of the FCC-ee injector study, we developed such a model that simulates $e^{+}$ production at the target and then tracks the resulting $e^{+}$ beam through the entire injector up to the DR \cite{FS_FCC}. Its application to the FCC-ee $e^{+}$ source is detailed in \cite{Alharthi:2025gpf}.
 
In this work, we present the first experimental validation of a complete start-to-end simulation framework for $e^{+}$ source modeling along with benchmarking against 
the most widely used simulation codes. These studies are performed at the SuperKEKB $e^{+}$ source, currently the only high-intensity facility in operation, making it an ideal reference for our study.

This paper is organized as follows: Sec.~\ref{sec:pos_sources_concept} provides an overview of the standard layout of the $e^{+}$ source, highlighting its challenges and then detailing the development of our simulation model. In Sec.~\ref{sec:SKEKB}, we introduce the $e^{+}$ source at SuperKEKB and demonstrate the application of our model to its configuration. Sec.~\ref{sec:benchmarking} presents the model validation results based on experimental measurements taken at the SuperKEKB $e^{+}$ source. Lastly, in Sec.~\ref{sec:conclusion}, we summarize our findings and future outlook.

\section{\label{sec:pos_sources_concept}\texorpdfstring{$e^{+}$}{e+} Sources Conceptual Design and Modeling}

$e^{+}$ beams can be generated using various techniques depending on the purpose of the experiment and the intensity requirements. In HEP colliders, where high intensity is crucial, $e^{+}$ sources almost universally rely on the conventional scheme. In this approach, a high-energy $e^{-}$ beam is directed onto a high-Z amorphous target, producing bremsstrahlung radiation. These high-energy photons then interact with the target nuclei, leading to the creation of $e^{-}e^{+}$ pairs. In principle, the $e^{+}$ production rate is governed primarily by two factors: the power (i.e., beam intensity and the energy of the primary $e^{-}$) and the target characteristics (material and thickness). While these parameters determine the $e^{+}$ production rate, practical implementation in the collider introduces additional challenges. In particular, the target must withstand a significant thermo-mechanical load to meet the high-intensity requirements. 

As the primary $e^{-}$ traverses the target, it loses energy through collisional and radiation processes. The radiative losses, namely bremsstrahlung, are carried off by secondary photons, while the collisional losses manifest primarily as ionization, which is the primary source of heating the target. Target heating considerations include both the mean energy deposition (average power) and the instantaneous energy deposition per pulse. The former sets the requirement for the cooling, and the latter is known as the Peak Energy Deposition Density (PEDD) during a beam pulse. PEDD can create thermal gradients that induce mechanical stresses, potentially leading to target failure if the PEDD exceeds the limit set by the material’s tensile strength. The conventional scheme was used for the $e^{+}$ source at the SLAC Linear Collider (SLC), where a 33~GeV $e^{-}$ beam was directed onto a tungsten-rhenium target with a thickness of six $X_0$. The SLC $e^{+}$ source holds the record for the highest intensity, achieving an intensity of approximately $6 \times 10^{12} \, e^{+}/\text{s}$ \cite{Chaikovska_2022}. 
Operational experience at SLC has shown that to prevent target damage, the PEDD in a tungsten-rhenium target should not exceed 35 J/g \cite{NLC:2001aa}. A similar threshold is considered for pure tungsten targets. 

Various design solutions have been proposed to address target heating issues (e.g., thermal stress, material fatigue, etc.). For instance, a moving target spreads successive pulses over a larger surface to reduce cumulative heating and mechanical stress on any single location. At SLC $e^{+}$ source, a trolling target was implemented as one such solution \cite{Reuter:trolling}, while a rotating target is currently under investigation for the $e^{-}$-driven $e^{+}$ source at ILC \cite{omori2024}. Another design is based on a compound target that incorporates an aligned crystal to exploit channeling radiation and an amorphous converter for $e^{+}$ production \cite{Chehab1989}. Such an approach, known as a hybrid scheme, helps reduce the PEDD by splitting the energy deposition between two targets. The hybrid scheme is proposed as the baseline for CLIC $e^{+}$ source \cite{Zhao:2025fzn} and has also been investigated as a potential alternative for the FCC-ee $e^{+}$ source \cite{Alharthi:2025gpf}. 

Following the $e^{+}$ production, an immediate capture system is essential since the generated $e^{+}$ beam typically experiences a large emittance and a significant energy spread due to multiple scattering. The initial capturing stage employs a Matching Device (MD) to adapt the beam for further transport. Common MD choices include an Adiabatic Matching Device (AMD) or a Quarter Wave Transformer (QWT) \cite{chehab_positron_1989}. The MD is primarily used to match the $e^{+}$ beam transverse profile to the subsequent sections. The choice of the MD is determined by the operational requirements of the $e^{+}$ source. An AMD is preferable when large energy and radial acceptance are required, while a QWT is better suited for cases that demand wide angular acceptance or the selection of specific energy bands. 

The subsequent $e^{+}$ source stage involves capturing the $e^{+}$ beam in Radio Frequency (RF) buckets of multiple accelerating structures in the longitudinal plane. Due to the large transverse emittance of the $e^{+}$ beam, the RF structures are immersed in a strong axial magnetic field (generated by DC solenoids) to focus the $e^{+}$ transversally. The combination of the MD and the first few accelerating structures is known as the capture section. At the end of the capture section, typically at a few hundred MeV,  a chicane with a collimator is used to intercept the secondary $e^{-}$. The $e^{+}$ are then further accelerated in an $e^{+}$ linac to reach the desired energy for the DR. 

\begin{figure}[h!]
\includegraphics[width=\columnwidth]{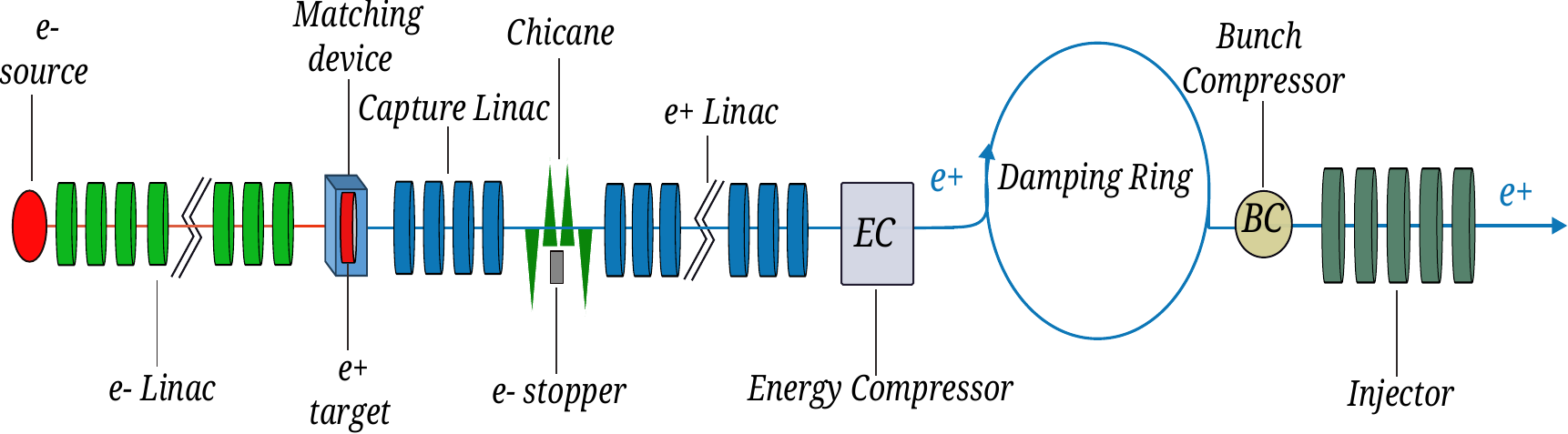}
\caption{\label{fig:ps_all}Schematic layout of different sub-systems of the $e^{+}$ source basic scheme.}
\end{figure}

A typical layout of the $e^{+}$ injector is presented in Fig.~\ref{fig:ps_all}. All these elements of the $e^{+}$ injector must be carefully optimized to maximize the accepted yield. Thus, developing a comprehensive start-to-end simulation tool is essential for achieving the ultimate performance of the injector and fulfilling the collider's requirements. 

The designing and modeling of $e^{+}$ sources are highly intertwined, with each stage of the process relying on specialized tools.
 For example, the initial modeling of pair production and the deposited energy in the target is typically performed using Monte Carlo (MC) codes such as \texttt{GEANT4} \cite{agostinelli2003geant4, allison2006geant4, allison2016recent}, \texttt{FLUKA} \cite{fluka1, fluka2}, or \texttt{EGS} \cite{EGS5}, which simulate particle interactions and radiative processes. Following this, the target thermo-mechanical aspects are simulated by \texttt{CST} \cite{CSTStudioSuite} or \texttt{ANSYS}~\cite{ansys}.
Once the $e^{+}$ are produced, beam dynamics codes like \texttt{ASTRA} \cite{astra}, \texttt{Elegant} \cite{Elegent}, \texttt{MAD-X} \cite{MadX}, \texttt{SAD} \cite{sad} , \texttt{GPT}, or \texttt{RF-Track}~\cite{RFT} are employed to simulate their capture, acceleration, and transport through the injector. It is important to note that changes in one part of the design, such as the target geometry, beam optics, or the DR acceptance, often mandate a corresponding change in another part. 
Therefore, iterative feedback between these simulation tools is essential.

This paper focuses on the model developed for the FCC-ee $e^{+}$ source, emphasizing both the $e^{+}$ production stage and the subsequent beam dynamics in the capture linac. Our start-to-end model integrates two main simulation tools. The widely known MC code \texttt{Geant4} version 11.2.2 utilizes the FTFP\_BERT physics list. This physics list offers a consistent description of electromagnetic and hadronic nuclear interactions over a broad energy range. Its accurate modeling of bremsstrahlung and pair production makes it well-suited for simulating $e^{+}$ generation in dense, high-Z materials. Subsequently, \texttt{RF-Track} code version 2.3.2 is employed for the following beam dynamics simulations in the capture section and the $e^{+}$ linac. \texttt{RF-Track} enables particle tracking in space and tracking in time. In our model, the equations of motion are integrated using time as the independent variable, solved numerically with the Runge-Kutta method. 
A convergence study was conducted to balance computational time and numerical accuracy. This analysis identified an integration step of 0.25~mm/c as the optimal choice.

\section{\texorpdfstring{Experimental validation of the model using the $e^{+}$ source at SuperKEKB}{Experimental validation of the model using the e+ source at SuperKEKB}}\label{sec:SKEKB}

The SuperKEKB facility is designed to collide $e^{-}$ and $e^{+}$ at the center-of-mass energies in the regime of B-meson, aiming to search for new physics beyond the SM \cite{SKEKB}. The facility comprises an injector linac (LINAC) equipped with $e^{-}$ and $e^{+}$ sources, an $e^{+}$ beam DR, a low energy ring (LER) for $e^{+}$ and a high energy ring for $e^{-}$, and the Belle II physics detector. The designed beam energies for the two rings LER and HER are 4~GeV and 7~GeV, respectively \cite{SKEKB}. 
As the successor to KEKB, which was in operation from 1998 to 2010 and
achieved the world-highest peak luminosity of 
$2.018 \times 10^{34}~cm^{-2}~s^{-1}$, SuperKEKB is upgraded to reach $8 \times 10^{35}~cm^{-2}~s^{-1}$,  ${40} \times$ KEKB luminosity \cite{Ohnishi2023}. 
At the end of the physics run in 2024, the achieved luminosity was $5.1 \times 10^{34}~cm^{-2}~s^{-1}$ \cite{KEKNews2025}, 2.5 times higher than the achieved luminosity at the KEKB. One of the primary key factors for achieving such a luminosity is the extensive upgrade and remodeling of the LINAC and the $e^{+}$ source.

KEK's LINAC has established a concurrent top-up injection to feed four storage rings. The LINAC is composed of eight sectors (sectors A-C and 1-5) with a bending sector in between (J-arc), as illustrated in Fig.~\ref{fig:KEK_injector}. Each sector has eight klystrons, with one klystron driving four 2~-m-long accelerating structures. The LINAC supplies an $e^{+}$ beam to the LER via $e^{+}$ DR, $e^{-}$ beam to the HER, and $e^{-}$ beams to two light sources: Photon Factory Advanced Ring (PF-AR) and Photon Factory storage ring (PF-ring) \cite{Natsui:2023jlc}. 
This simultaneous injection is maintained by utilizing two independent $e^{-}$ sources operating up to 50~Hz, with each RF pulse delivering two bunches to both the HER and the LER.
One source \textemdash is a DC thermionic gun \textemdash supplies the $e^{-}$ beam for the two light sources and the $e^{+}$ target providing charges of 0.3~nC and 10~nC respectively.
The other source, \textemdash a photocathode RF gun \textemdash generates a low-emittance $e^{-}$ beam specifically for the HER. This paper will focus mainly on the $e^{+}$ source and the following capture linac located in Sector~1 within the LINAC.

\subsection{\texorpdfstring{$e^{+}$}{e+} source at the SuperKEKB}

The $e^{+}$ source at SuperKEKB represents the current state-of-the-art and is the world's highest-intensity $e^{+}$ source in operation \cite{Kamitani2014}. It employs the conventional scheme, where a primary $e^{-}$ beam (generated by a thermionic gun with an intensity of 10~nC per bunch) with an energy of 2.9~GeV impinges upon a tungsten target. 

\onecolumngrid

\begin{center}
\includegraphics[width=0.85\textwidth]{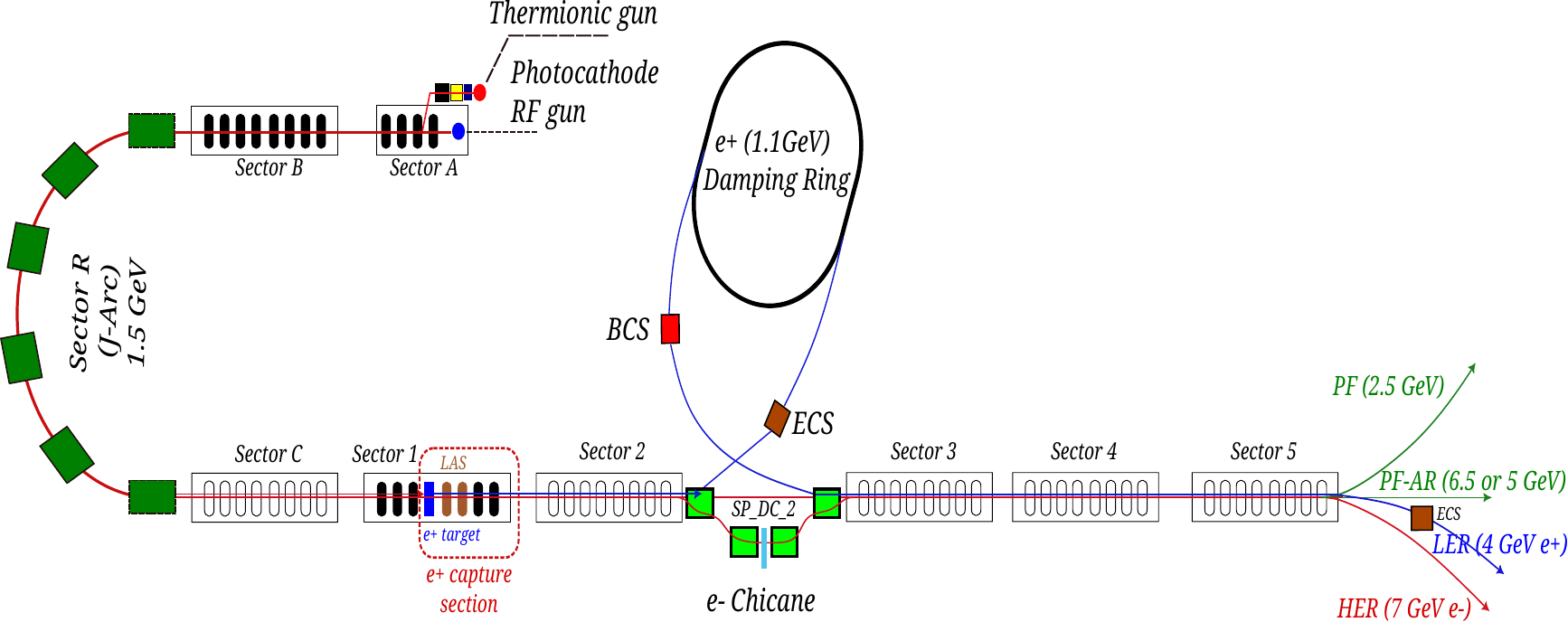}
\captionof{figure}{\label{fig:KEK_injector} Schematic layout of the KEK injector LINAC.}
\end{center}

\twocolumngrid

\begin{figure}[H]
\includegraphics[width=0.95\columnwidth]{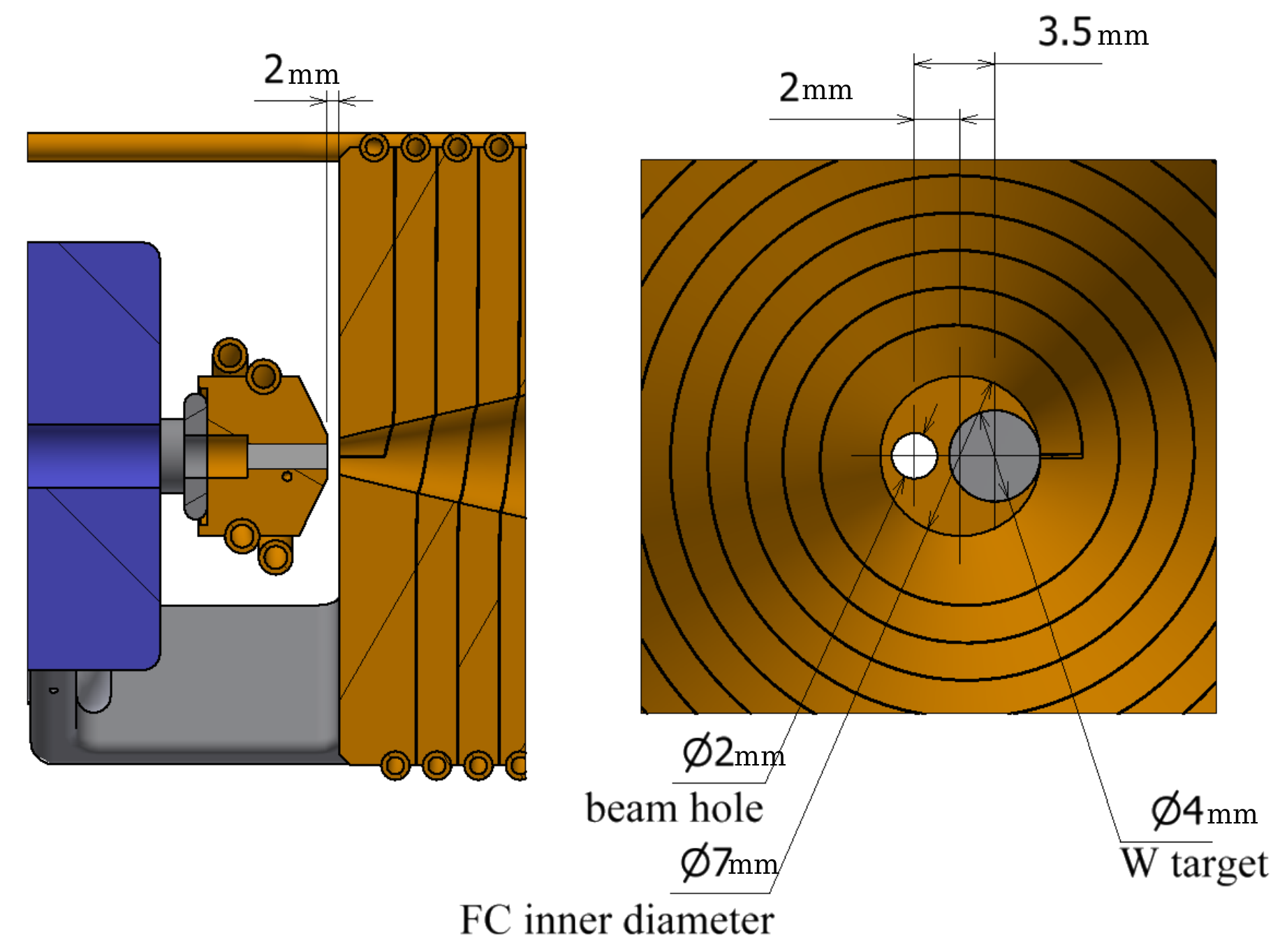}
\caption{\label{fig:target} Schematic layout of the SuperKEKB $e^{+}$
source target: side view (left) and downstream view (right), illustrating the target offset position. The beam hole center serves as the reference point. }
\end{figure}

The target is a 14~mm-thick (4~$X_0$) cylinder with a 4~mm diameter, and it is housed in a copper holder to ensure efficient cooling and support.
Since the LINAC is shared between $e^{-}$ and $e^{+}$ bunches, the $e^{+}$ target's center is displaced by 3.5~mm from the beam axis, and a hole with a 2~mm diameter is bored in the copper along the beam axis for $e^{-}$ beam passage to the HER \cite{Zang2014}, as shown in Fig.~\ref{fig:target}.

\begin{figure}[h!]
\includegraphics[width=\columnwidth]{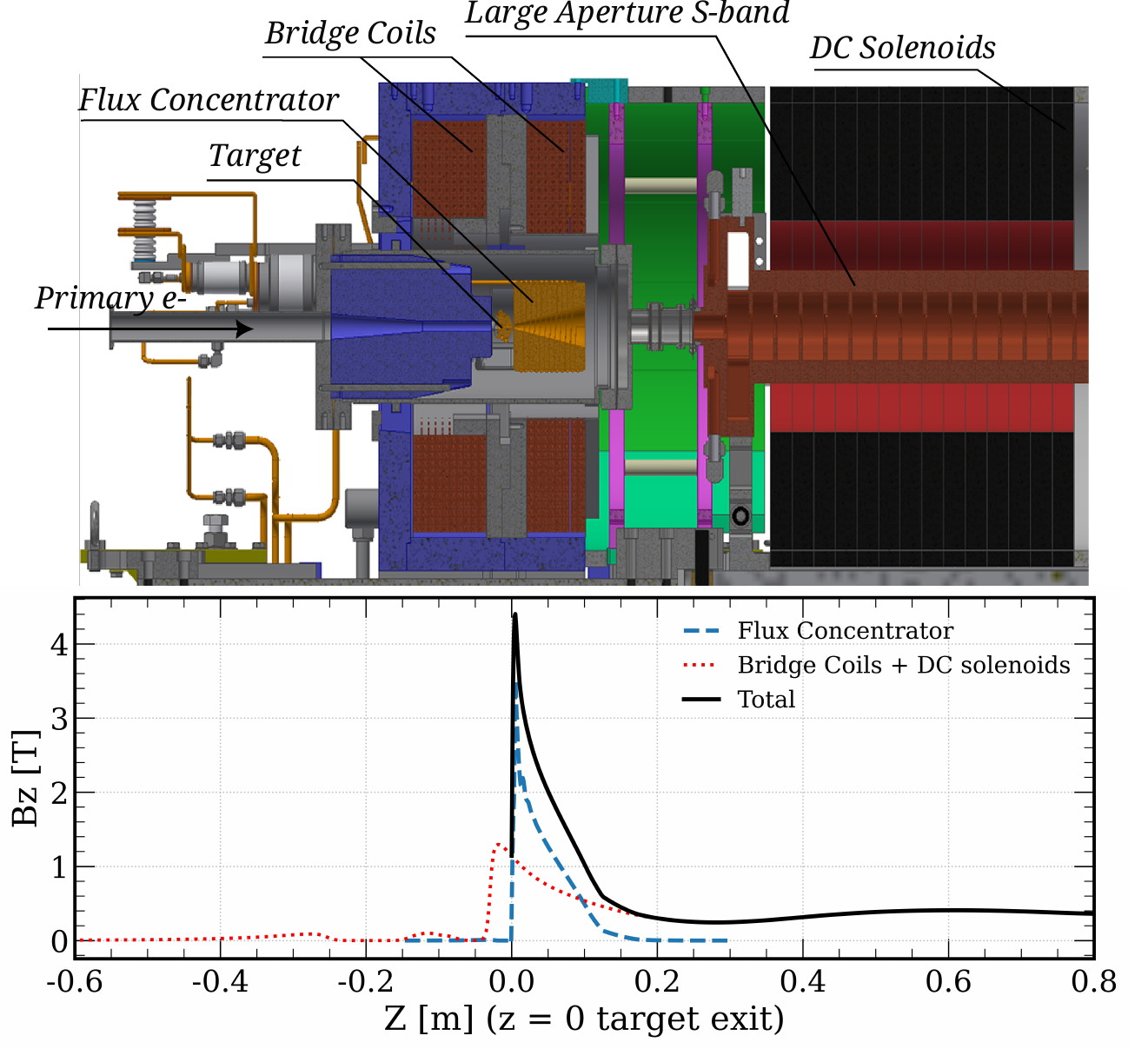}
\caption{\label{fig:AMD} Cross-sectional view of the SuperKEKB 
$e^{+}$ source target, displaying the flux concentrator, bridge coils, and the DC solenoids with the corresponding longitudinal magnetic field profiles.}
\end{figure}

Upstream of the target, a series of pulsed magnets are positioned to steer the primary $e^{-}$ beam off-axis, directing it to hit the target. Downstream of the target, an MD in the form of a Flux Concentrator (FC) is installed to increase the $e^{+}$ capture efficiency. The FC is a pulsed solenoid magnet made of copper alloy featuring an internal truncated cone shape, with front and rear aperture diameters of 7~mm and 52~mm, respectively \cite{Enomoto2021}. The total length of the FC is 100~mm, and its center is positioned at x = 2~mm relative to the beam axis to accommodate the target offset.
The strong magnetic field inside the FC is excited by a few microseconds pulse with a peak current of 12~kA.
As shown in Fig.~\ref{fig:AMD}, the longitudinal magnetic field profile rises sharply to a peak value of 3.5~T at the FC entrance, then gradually tapers down to 0.4~T, matching the field of the downstream DC solenoids surrounding the RF structures. DC solenoids, the so-called Bridge Coils (BC), have been added to enhance the FC field and improve capture efficiency. The FC has been operating since 2020 without any significant issues. A detailed view of the target area is presented in Fig.~\ref{fig:AMD}. 

The distance between the FC exit and the RF structures is minimized to 137~mm to preserve the quality of the $e^{+}$ beam. The capture linac comprises six Large Aperture S-band (LAS) traveling wave accelerating structures, whose parameters are listed in Table~\ref{tab:RF_parameters}.

\begin{table}[h!t]
\caption{\label{tab:RF_parameters}Parameters of the RF structures in the capture linac.}
\begin{ruledtabular}
\begin{tabular}{@{}lcc@{}}
\textbf{Parameter}   & \textbf{Value}            & \textbf{Units}  \\
Type  &  LAS              \\
Frequency        & 2.856 &          GHz         \\
Phase advance        & 2$\pi$/3 &          deg         \\
Number of cells     & 59                   \\
Iris diameter     &  31.861 -- 29.967 &  mm              \\
Length        &  2.064 &          m         \\
Number of LAS       &  2 + 4 &                   \\
Klystron power no--SLED        & 40 &          MW         \\
SLED Gradient (Max)        &  14/10  &          MV/m        \\
\end{tabular}
\end{ruledtabular}
\end{table}
Initially, the RF units consisted of a 40~MW klystron and an SLED that fed four LAS \cite{Matsumoto2014}. However, to have a flexible setting of the RF gradient in the upstream part of the capture linac, the RF unit KL15 drives only two LAS, while the KL16 RF unit powers the subsequent four structures.
At the solenoid channel's exit, located at the end of the last RF structure, several focusing and defocusing quadrupoles are installed to match the beam to the following acceleration sections. Downstream of these quadrupoles, a chicane system is implemented with a peak field of 0.2~T. An $e^{-}$ stopper is placed at the chicane's center to intercept the secondary $e^{-}$. Additional quadrupoles are installed after the chicane to complete the beam-matching to the following FODO lattices focusing-based accelerating sections.

\begin{figure}[h!]
\centering
\includegraphics[width=\columnwidth]{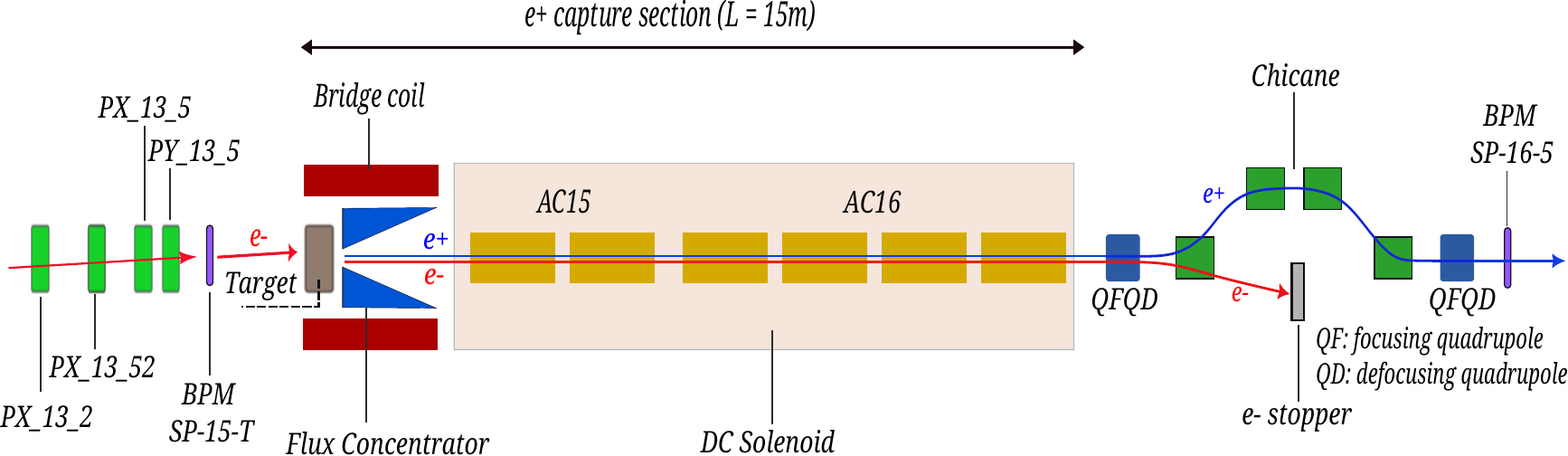}
\caption{\label{fig:CS} Schematic layout of the SuperKEKB $e^{+}$ source.}
\end{figure}

Due to the compactness of the capture section and the extremely short time interval between the $e^{+}$ and secondary $e^{-}$ bunches (typically on the order of 100~ps), the use of conventional beam diagnostics within this region is highly constrained. Therefore, in the current setup, only two conventional Beam Position Monitors (BPMs) are employed for beam diagnostics.  The first BPM, labeled as \text{SP\_15\_T}, is positioned just before the target and is used for primary $e^{-}$ beam diagnostics. The second BPM, labeled \text{SP\_16\_5}, is located after the chicane and used for $e^{+}$ beam diagnostics. The schematic layout of the SuperKEKB $e^{+}$ source is depicted in Fig.~\ref{fig:CS}.

\begin{figure}[!htb]
\includegraphics[width=\columnwidth]{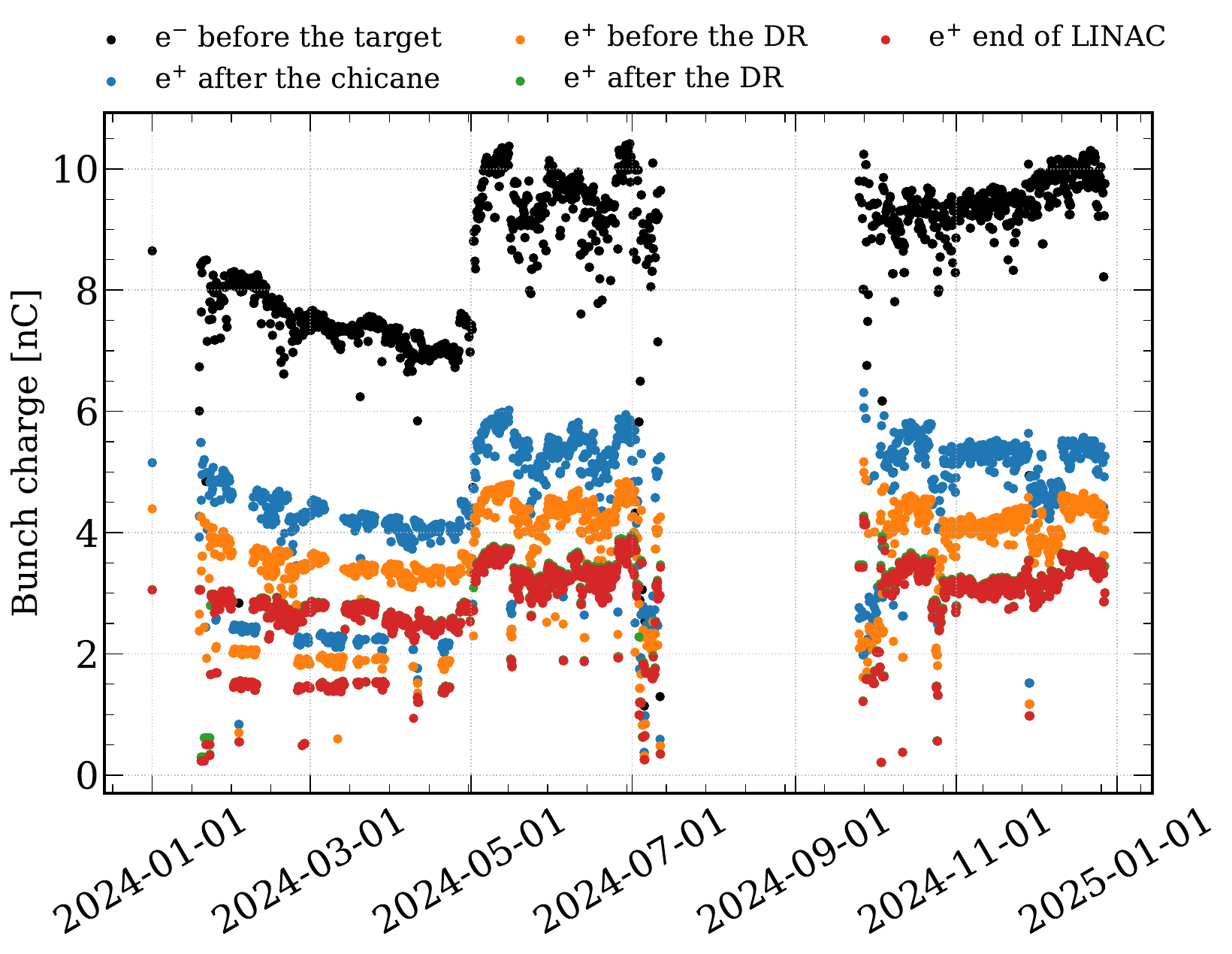}
\caption{\label{fig:bpm} Bunch charge evolution throughout the $e^{+}$ production and acceleration process during the 2024 run. The primary $e^{-}$ bunch charge before the  $e^{+}$ target is shown in black. The $e^{+}$ bunch charge is plotted at different stages: after the target at the end of the capture section (blue), before injection into the DR (orange), at the DR exit (green), and at the linac exit (red). The gap represents the summer shutdown.}
\end{figure}

The bunch charge history from the 2024 run measured at different locations along the LINAC is shown in Fig.~\ref{fig:bpm}. The primary $e^{-}$ bunch typically had a charge of 10~nC. The $e^{+}$ bunch charge, which was 6~nC after the chicane, gradually decreased to 4~nC at the DR entrance, corresponding to an $e^{+}$ yield of 0.4. Beyond the DR and through the remaining part of the LINAC, the $e^{+}$ bunch charge stabilized at 3.5~nC.

\subsection{$e^{+}$ production and tracking}

We applied the model mentioned in the previous section to the SuperKEKB  $e^{+}$ source. The simulation process consists of the following steps: first, a primary $e^{-}$ beam is generated based on the parameters listed in Table~\ref{tab:CS_parameters}. Second, the beam is tracked through the steering magnet in the upstream section of the target using \texttt{RF-Track}. Third, the tracked $e^{-}$ particles are imported into \texttt{Geant4}, where the $e^{+}$ production process is simulated. Finally, the resulting $e^{+}$ bunch is extracted and re-imported into \texttt{RF-Track} for further beam dynamics studies through the capture section.

\begin{figure}[h!]
\centering
\includegraphics[width=0.9\columnwidth, keepaspectratio]{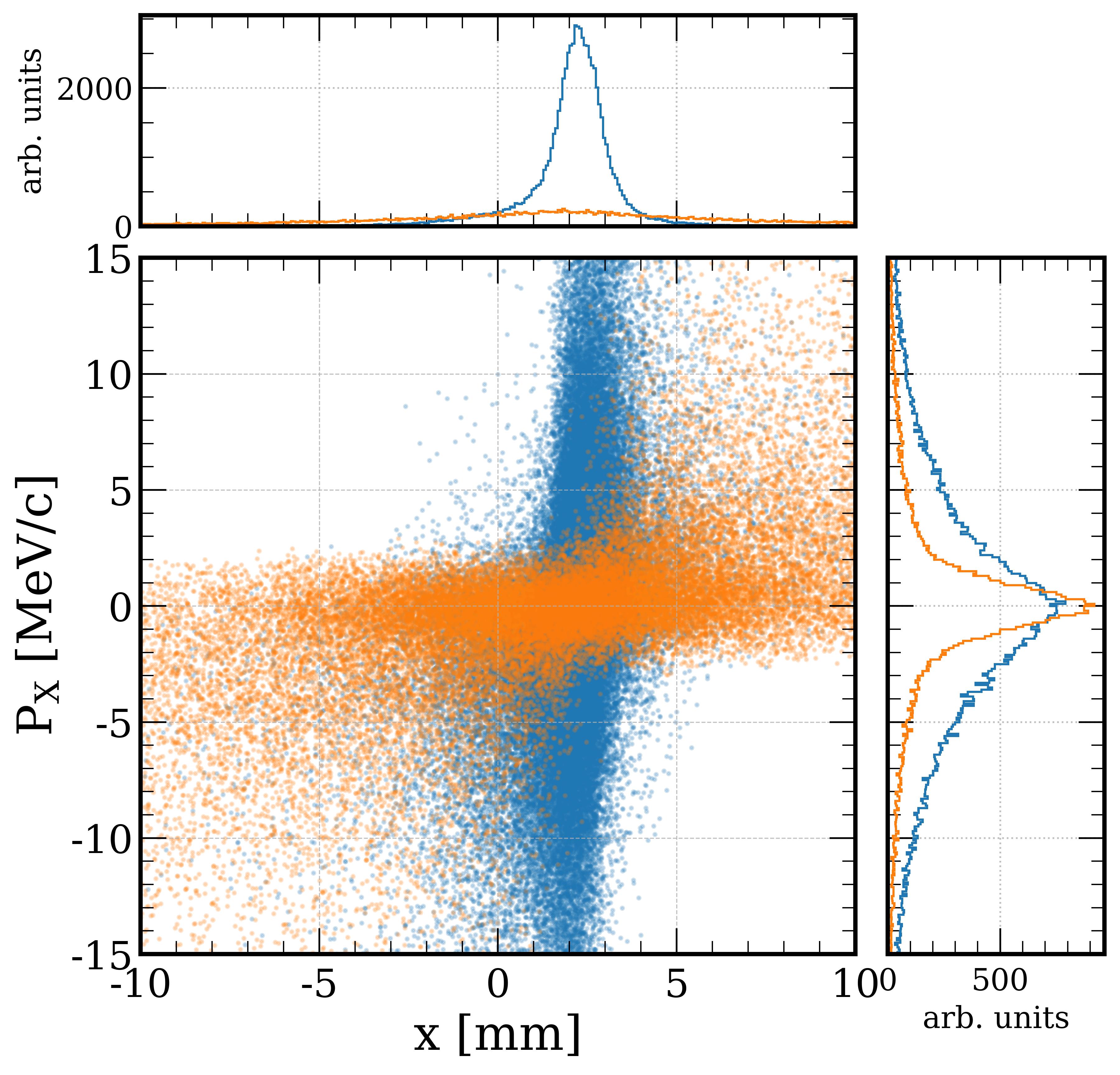}
\caption{\label{fig:Transverse} Transverse phase space of the $e^{+}$ at the target exit (blue) and at the exit of the FC (orange).}
\end{figure}

The flexibility of the developed model in \texttt{Geant4} allows for a precise definition of incoming beam properties, including energy, emittance, energy spread, and time structure. Furthermore, the model considers the full design of the target, including the copper holder, the tungsten target, and the hole for $e^{-}$ beam passage. According to \texttt{Geant4} simulation performed based on the parameters listed in Table~\ref{tab:CS_parameters}, an $e^{+}$ yield of 7.6 $N_{e^{+}}/N_{e^{-}}$ is emerged from the target exit side. As clearly shown in Fig.~\ref{fig:Transverse}, the $e^{+}$ produced has a large transverse momentum spread ($\sigma_{Px}$ = 4.33~MeV/c) and a small lateral size ($\sigma_{x}$ = 0.63~mm). Along the longitudinal direction, the phase space is non-Gaussian and characterized by a momentum spread of $\sigma_{P_z}$ = 97.1~MeV/c, with a mean around 44.7~MeV/c. These values indicate that $e^{+}$ dynamics are strongly influenced by the large transverse emittance and large energy spread.

\begin{table}[h!]
\caption{\label{tab:CS_parameters} SuperKEKB $e^{+}$ source operation conditions measured during the experimental campaign.}
\begin{ruledtabular}
\begin{tabular}{@{}lcc@{}}
\textbf{Parameter}   & \textbf{Value}            & \textbf{Units}  \\
Primary ${e^{-}}$ energy        & 2.9 &          GeV         \\
Repetition rate   & 25    & Hz \\
Primary ${e^{-}}$ charge    & 2$\times$10 & nC \\
Beam power & 1.45 & kW \\
Beam size $\sigma_x$/$\sigma_y$          & 0.45/0.52 & mm              \\
Position on the target x/y        & 2.25/0.2 &          mm                 \\
Estimated power deposited in the target & 0.41 & kW  \\
PEDD & 21 & J/g \\
Estimated gradient (KL15/KL16)     &  8.13/9.47 &  MV/m              \\
${e^{+}}$ charge after the chicane       &  6  &    nC        \\
${e^{+}}$ yield after the chicane        &  0.6 &    ~$N_{e^{+}}/N_{e^{-}}$               \\

\end{tabular}
\end{ruledtabular}
\end{table}

\begin{figure}[h!]
\includegraphics[width=\columnwidth]{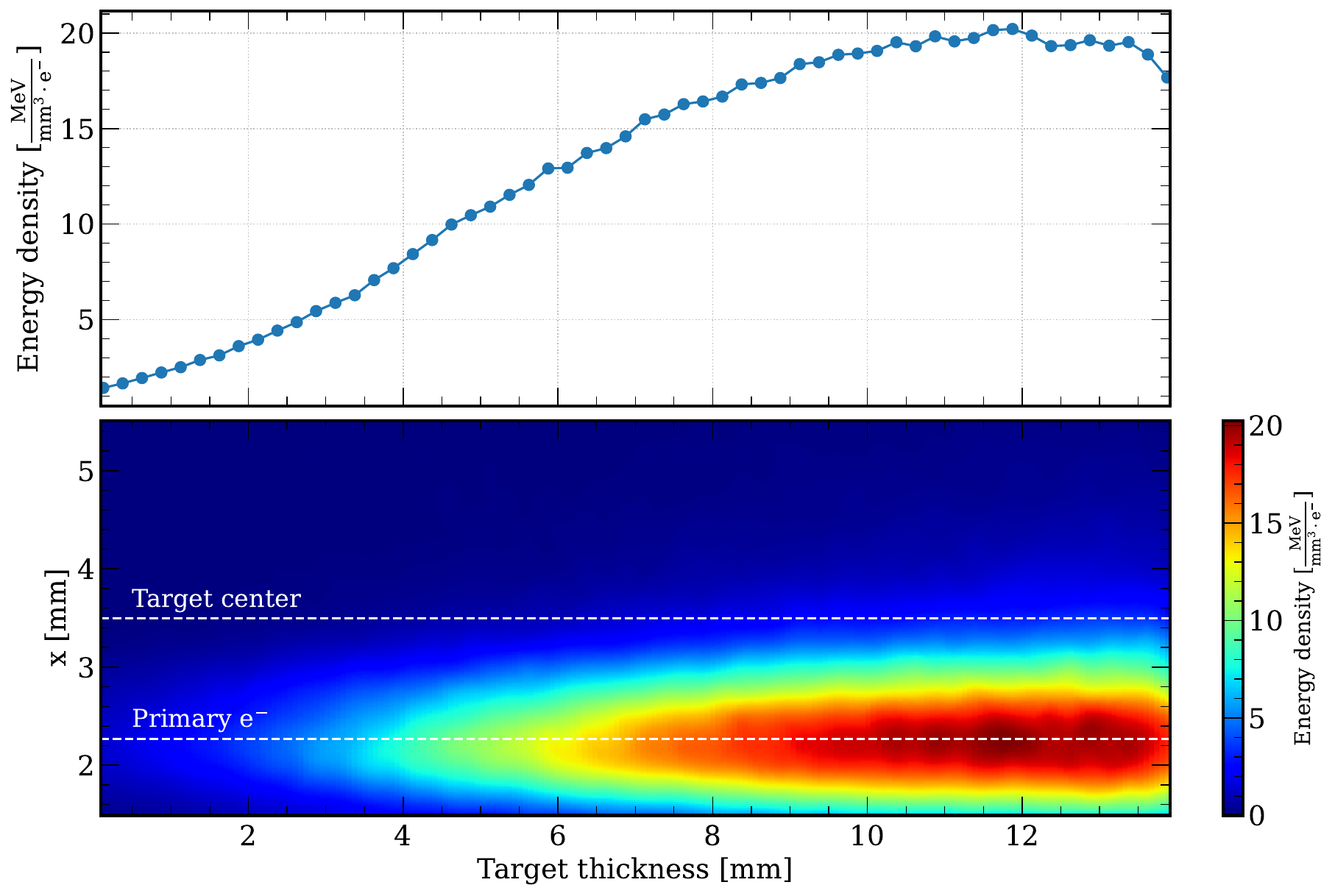}
\caption{\label{fig:PEDD_G4} Energy density along the primary $e^{-}$ beam axis in the tungsten target, with a maximum value of \(21~\text{MeV/mm}^3\) per incident $e^{-}$ observed near the target exit (top), and electromagnetic shower development within the target (bottom).}
\end{figure}

In addition to the $e^{+}$ production, the model assesses the integral energy deposited in the target per incident $e^{-}$, which can be scaled according to the primary $e^{-}$ bunch charge. This parameter helps to estimate the target heating and, therefore, sets the cooling requirements. Also, the model can estimate the PEDD in the target using a specific scoring function in \texttt{Geant4} (mesh scorer). This function defines a cubic voxel grid, where each voxel records the energy deposited per incident $e^{-}$. The voxel dimensions are set to half of the beam size  ($\Delta x = \Delta y = \Delta z = \frac{\sigma_x}{2})$ \cite{Zhao_PEDD}. The development of the electromagnetic shower within the target and the corresponding energy density distribution along the beam axis are presented in Fig.~\ref{fig:PEDD_G4}. Due to shower development within the target, the maximum energy density was found near the exit face, reaching  \(21~\text{MeV/mm}^3\) per incident $e^{-}$. Based on this value, the estimated PEDD per pulse is 21~J/g.

At the beginning of the capture section, the $e^{+}$ bunch passes through the FC, where it experiences a strong axial magnetic field that gradually decreases to match the solenoid field of the capture linac. 
This tapered magnetic field in the FC influences the beam dynamics in both the transverse and longitudinal planes. In the transverse plane, the gradual field variation induces a phase-space rotation, which reduces the transverse momentum spread to $\sigma_{px}$ = 1.37~MeV/c while maintaining an acceptable beam size $\sigma_x$ = 4.63~mm (capture linac aperture 2a$=$30~mm), as shown in Fig.~\ref{fig:Transverse}. In the longitudinal plane, $e^{+}$ follow helical trajectories due to the axial magnetic field. The transverse momentum governs the radius of this helical motion: $e^{+}$ with higher transverse momentum follow wider spirals, while those with lower transverse momentum remain in tighter spirals. This momentum-dependent trajectory variation leads to bunch lengthening as the $e^{+}$ bunch exits the FC. As a result, the longitudinal capture efficiency may decrease, since differences in path length spread the bunch longitudinally, making it harder to focus in the capture linac.
 
In the capture linac, the $e^{+}$ are longitudinally captured within RF buckets. The electromagnetic field map is generated using the SuperFish code \cite{superfish}. It is then imported into \texttt{RF-Track} and interpolated using the Cubic Interpolation (CINT) function. This method evaluates the RF field at any given point by performing a cubic interpolation over the nearest 64 mesh points, ensuring a smooth and accurate field representation. The first LAS structure should operate in deceleration mode (deceleration RF phase) to enhance bunching efficiency and maximize the number of $e^{+}$ confined within a single RF bucket.

\begin{figure}[h!]
\centering
\includegraphics[width=\columnwidth, keepaspectratio]{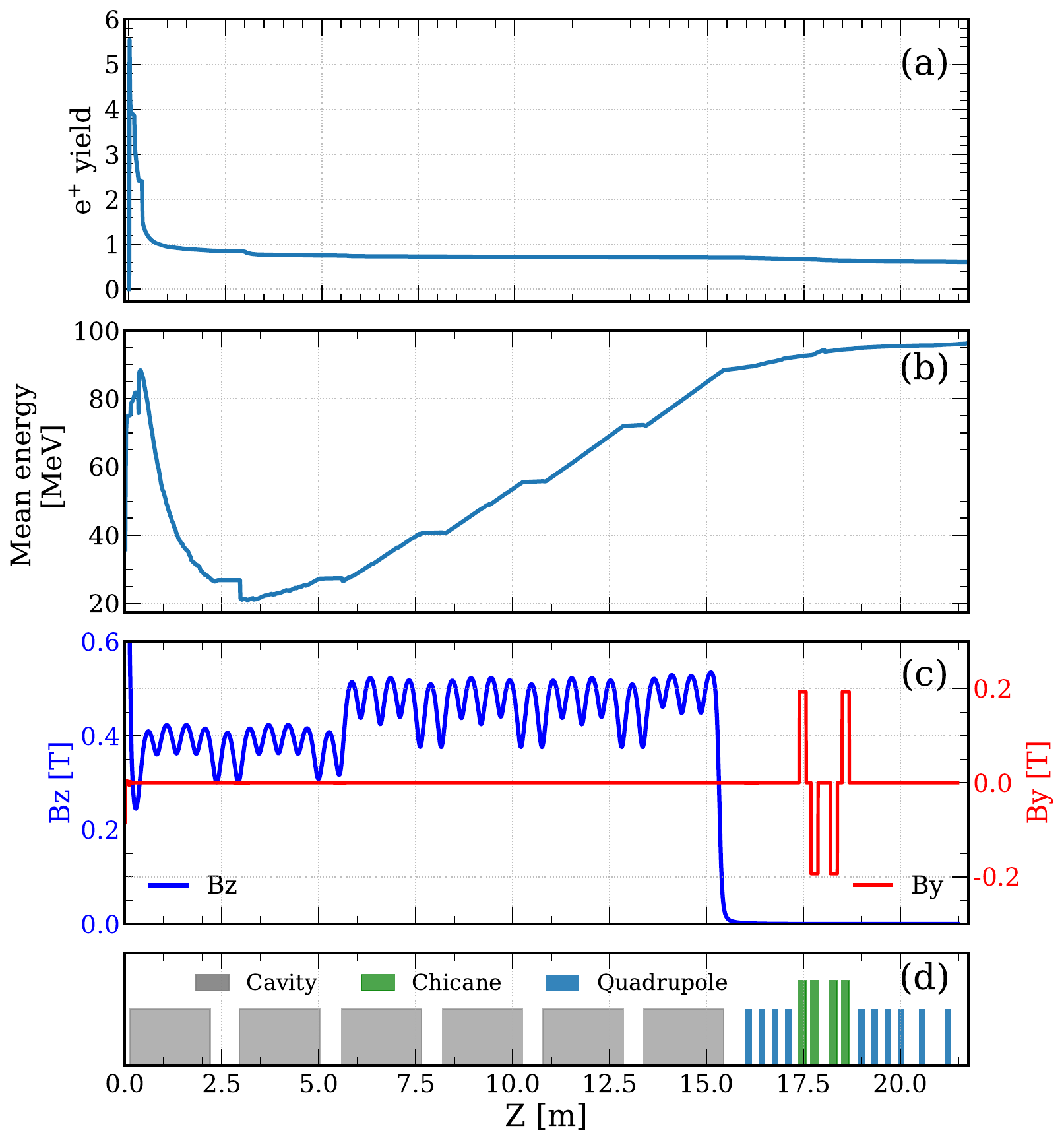}
\caption{\label{fig:TT}Simulation results of the SuperKEKB $e^{+}$ capture section, illustrating yield evolution (a), mean energy gain (b), magnetic field profiles (c), and schematics of the section's elements (d). At the end of the capture section, an $e^{+}$ yield of 0.6 and a mean energy of 100~MeV were found.}
\end{figure}

According to the simulation results shown in Fig.~\ref{fig:TT}, an $e^{+}$ yield of 0.60 was observed at the end of the capture section. This yield is in excellent agreement with the charge history shown in Fig.~\ref{fig:bpm}. A significant portion of $e^{+}$ losses, approximately 70\%, occurs between the target exit and the entrance of the first RF structure, primarily due to its aperture. 
Further losses, approximately 5\%, were observed in the chicane, where low-energy $e^{+}$ were lost upon striking the edge of the beam pipe at the chicane center. Additionally, at the beginning of the capture linac, $e^{+}$  are decelerated to enhance the bunching and reduce the energy spread, ultimately improving the accepted $e^{+}$ yield. At the end of the capture section, the $e^{+}$ mean energy is approximately 100 MeV.

\section{Benchmarking Results and Discussion}\label{sec:benchmarking}

To assess the reliability of our model, we validated its results with experimental measurements conducted at the SuperKEKB $e^{+}$ source and benchmarked it using independent simulation tools. These tools can be categorized into two main approaches. The first approach corresponds to the model used at KEK, where \texttt{EGS5} is employed for $e^{+}$ production, integrated with \texttt{GPT} for particle tracking. For completeness, the second approach utilizes the same \texttt{Geant4} model presented earlier; however, the tracking is performed using \texttt{ASTRA} instead of \texttt{RF-Track}.

We performed three key measurements to evaluate the accuracy and reliability of the presented model. Corresponding simulations were conducted for each measurement to compare with the experimental data. In each measurement, we scan one parameter while observing two BPMs: SP\_15\_T to measure the position and charge of the primary $e^{-}$, and SP\_16\_5 to measure the $e^{+}$ charge after the chicane (see Fig.~\ref{fig:CS}). The Figure of Merit (\textbf{FoM}) in all the measurements is the $e^{+}$ yield after the chicane

\begin{equation}
\text{FoM} = \frac{Q_{e^{+}} \text{ at SP\_16\_5} }{Q_{e^{-}} \text{ at SP\_15\_T}}
\end{equation}

\begin{figure}[!htb]
\includegraphics[width=.98\columnwidth]{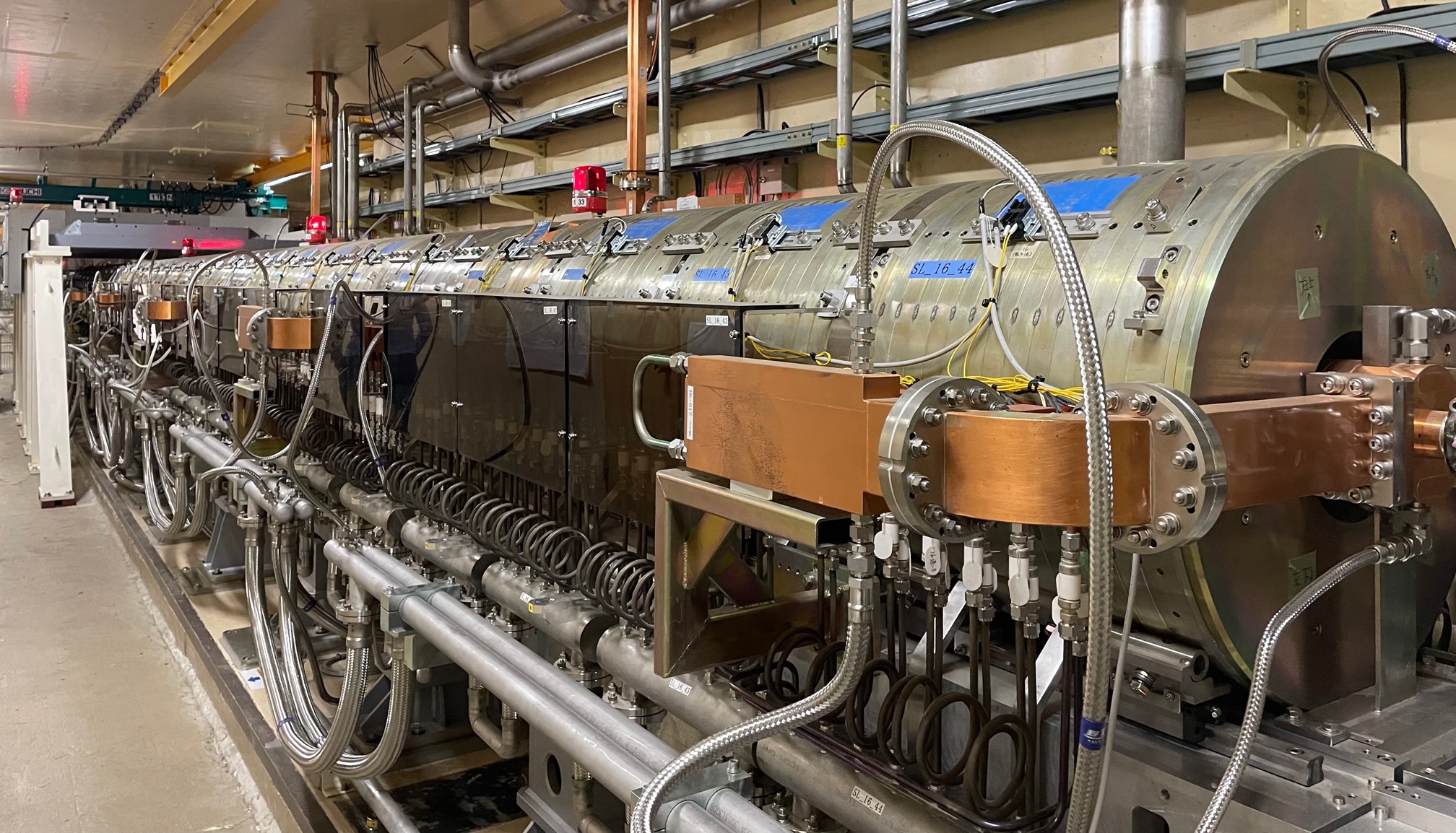}
\caption{\label{fig:captureLinac}SuperKEKB $e^{+}$ capture section. Photographed in November 2023 during the maintenance day.}
\end{figure}

Before each scan, we measure the primary $e^-$ transverse size using an $\mathrm{Al}_2\mathrm{O}_3:\mathrm{Cr}$
screen located 1~m upstream of the target. The beam size remained stable throughout the measurement campaign at approximately $\sigma_x = 0.45$~mm and $\sigma_y = 0.52$~mm. The actual SuperKEKB $e^{+}$ capture section where the measurements took place is shown in Fig.~\ref{fig:captureLinac}

\subsection{Field scan of the solenoids in the capture section}

The DC solenoids surrounding the accelerating structures in the capture linac are critical in focusing the $e^{+}$ beam, as discussed in Sec.~\ref{sec:pos_sources_concept}. The magnetic field profile varies along the capture linac, starting at approximately 0.4~T before increasing to 0.5~T, as shown in Fig.~\ref{fig:TT}. This variation arises from historical reasons, as several injector components were repurposed from KEKB. Additionally, the field is not perfectly axisymmetric, causing a transverse kick to the $e^{+}$ beam, reducing beam transport efficiency. Four steering magnets were installed around the capture linac to compensate for this effect. In this measurement, the magnetic field of the BCs,  the DC solenoids, and the steering magnets were gradually scanned from 100\% to 50\% in 10\% increments.

\begin{figure}[h!]
\centering
\includegraphics[width=\columnwidth]{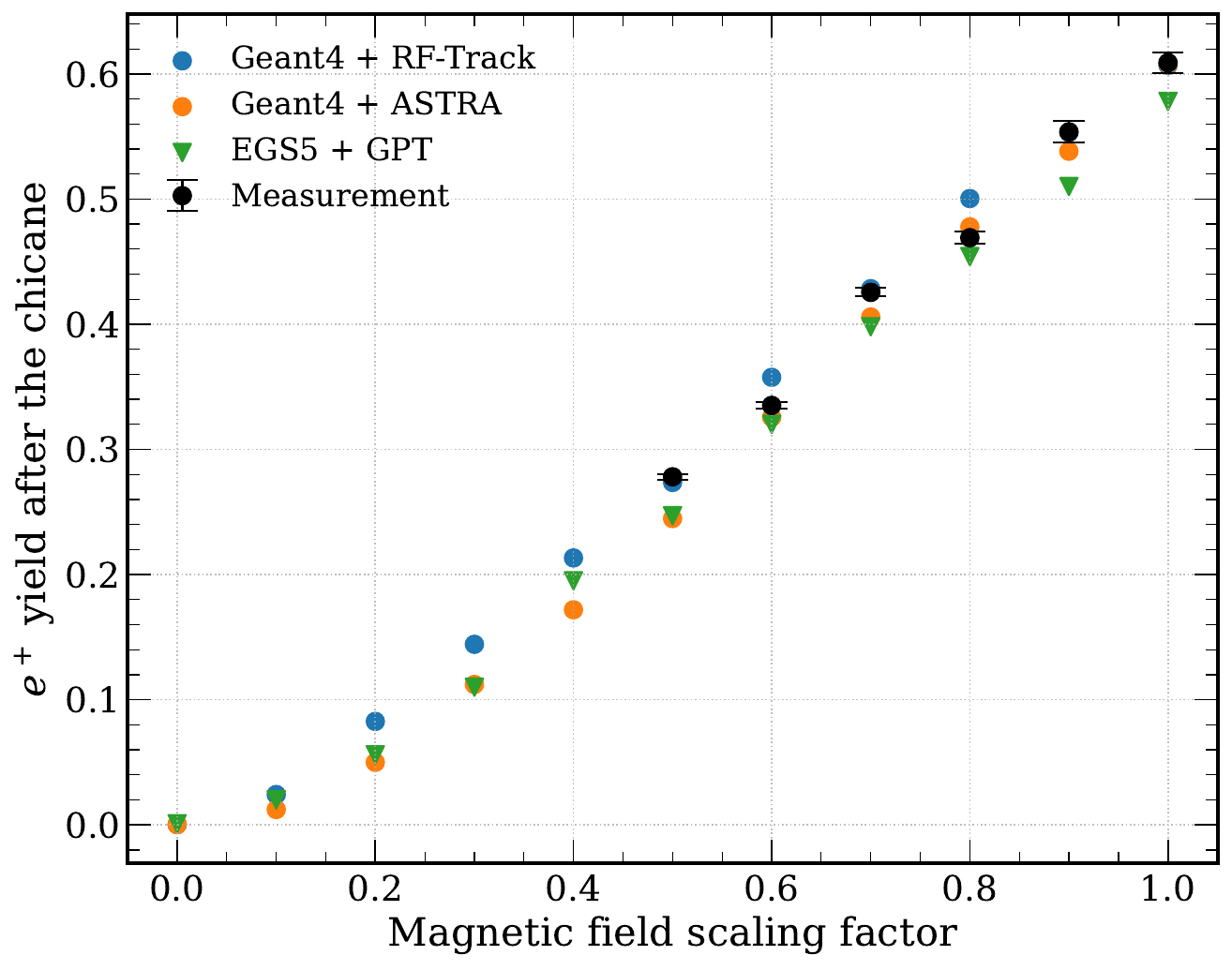}
\caption{\label{fig:DC} $e^{+}$ yield as a function of the magnetic field of the BCs, DC solenoids, and the steering magnet in the capture linac. Each measurement point was averaged over 50 acquisitions to improve statistical accuracy. Error bars in the figure indicate the resulting standard deviations.}
\end{figure}

The measurement results within the scan range shown in Fig.~\ref{fig:DC} reveal a linear relationship between the FoM and the magnetic field strength, highlighting the role of the DC solenoids in $e^{+}$ capture efficiency. A corresponding simulation was performed and extended to a region without the magnetic field. The simulation results show good agreement with the measurements and confirm that almost no $e^{+}$ could reach the end of the capture section without the magnetic field of the DC solenoid channel.

\subsection{Impact position scan of the primary $e^-$ on the target}

This measurement involved scanning the primary $e^-$ beam impact position on the target, conducted horizontally and vertically. As described in Sec.~\ref{sec:SKEKB}, the target center is positioned 3.5~mm horizontally from the beam axis, and the FC center is located 2~mm horizontally from the beam axis. Two steering magnets are mainly used in the measurement to steer the beam upstream of the target. \( \text{PX\_13\_5} \) magnet for horizontal steering and  \( \text{PY\_13\_5} \) magnet for the vertical steering  of the primary $e^{-}$ (see Fig.~\ref{fig:CS}). This scan aimed to evaluate how the primary $e^{-}$  beam position affects the FoM and investigate the influence of beam alignment on $e^{+}$ production. 

\begin{figure}[h!]
\centering
\includegraphics[width=\columnwidth]{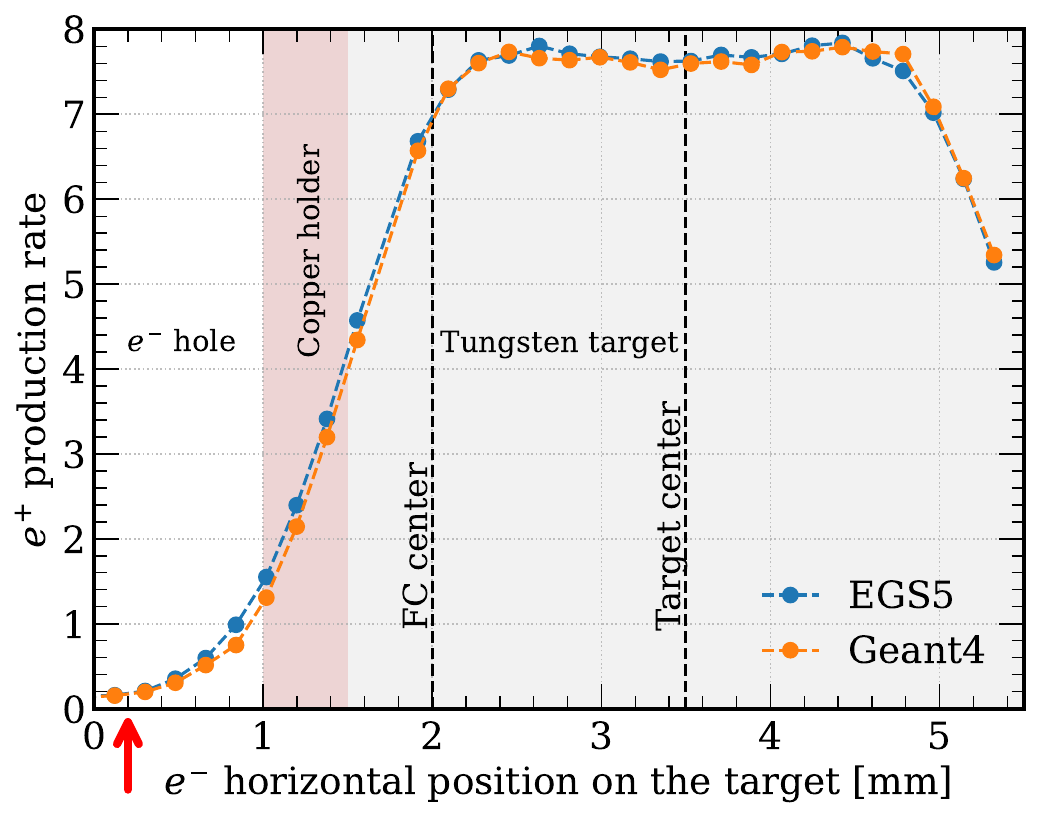}
\caption{\label{fig:G4_EGS5} Simulated $e^{+}$ production rate as a function of the primary $e^{-}$ horizontal impact position. The primary $e^{-}$  beam is orthogonal to the target, and its direction is indicated by the arrow in the lower left corner. The simulation data points are connected as a guide to the eye and do not represent a fit or model.}  
\end{figure}

Before presenting the measurement results, we will compare the two MC codes, \texttt{Geant4} and \texttt{EGS5}, in simulating the $e^{+}$ production after the target. The comparison was performed by moving the primary $e^{-}$ horizontally across the target, including its copper holder. The results demonstrate excellent agreement between the two simulation tools, as shown in Fig.~\ref{fig:G4_EGS5}. A similar level of agreement was observed when performing the primary $e^{-}$ vertical scan on the target.

\begin{figure}[h!]
\centering
\includegraphics[width=0.98\columnwidth]{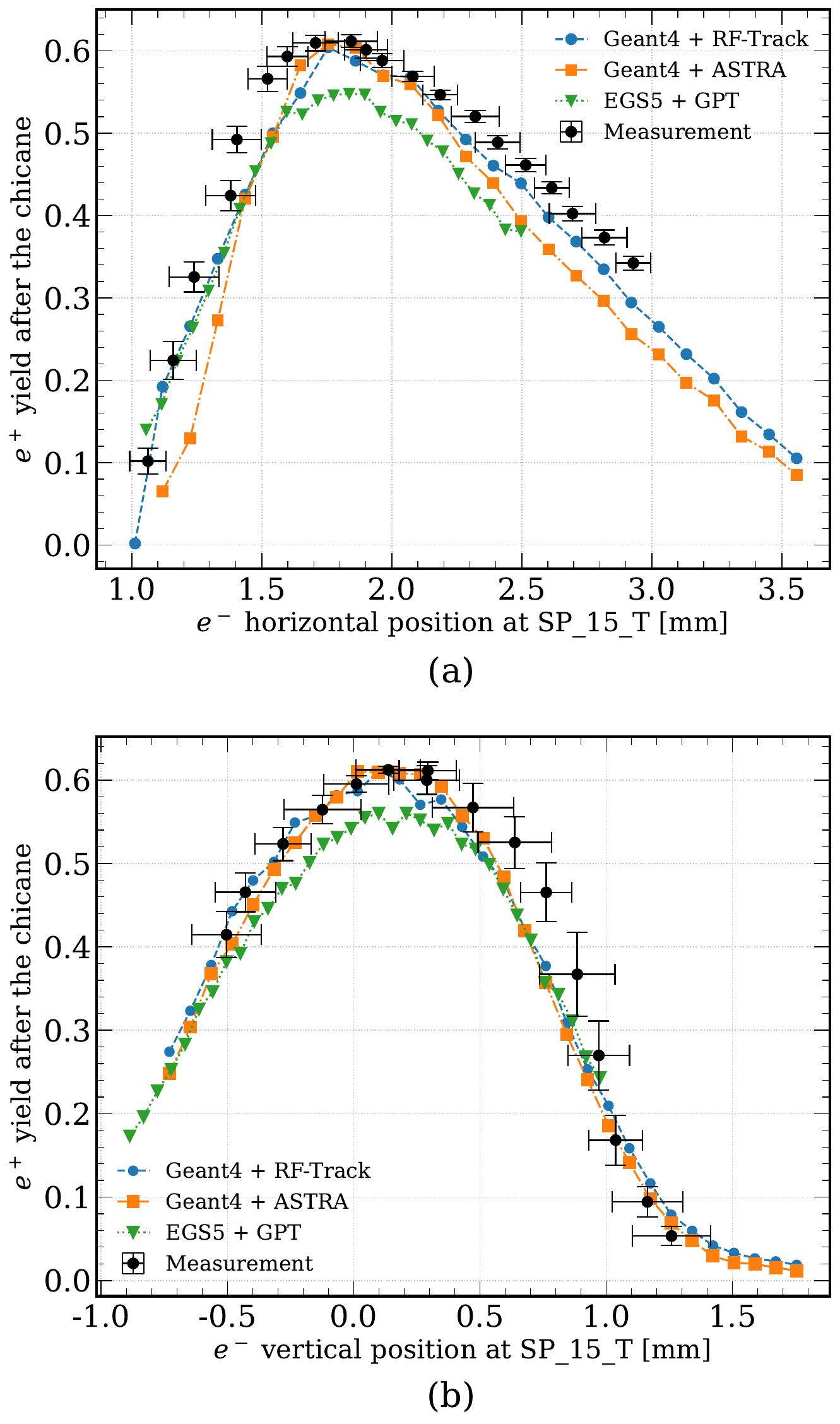}
\caption{\label{fig:hor} $e^{+}$ yield as a function of the primary $e^{-}$ beam impact position at the SP\_15\_T BPM just upstream of the target. (a) Horizontal impact position scan, (b) Vertical impact position scan. Each measurement point was averaged over 50 acquisitions to improve statistical accuracy. However, due to operational conditions on the measurement day, position jitter and charge jitter for the primary $e^{-}$ beam were observed and reported by the error bars in the figure. The simulation data points are connected as a guide to the eye and do not represent a fit or model.}
\end{figure}

The measurement and simulation results of the horizontal position scan of the primary $e^{-}$ on the target are shown in Fig.~\ref{fig:hor}(a). 
The comparison of the simulations and experimental data manifests a very good agreement. It is important to note that the primary $e^{-}$ beam is not perpendicular to the target face but arrives at an angle due to the trajectory deviation induced by the steering magnet. 

The actual impact position of the $e^{-}$ beam on the target is 2.25~mm, as determined from the simulated trajectory. Since the SP\_15\_T BPM is located 1.2~m upstream, it measures a beam position of 1.8~mm under nominal operating conditions, slightly offset due to the beam’s angular deviation. The $e^{+}$ yield profile remains nearly flat for horizontal positions between 1.5~mm and 2~mm, i.e., in the vicinity of the FC centre. 
This provides a wide horizontal impact position tolerance, almost equal to the beam size. However, a sharp drop in $e^{+}$ yield is observed below 1.5~mm, where $e^{+}$ originated from the copper holder rather than the target itself.

The next measurement involved a vertical scan of the primary $e^{-}$ beam impact position on the target using the same method. During the measurement, the $e^{-}$ beam position was scanned vertically while maintaining the optimal horizontal position at 2.25 mm. The measurement and simulation results of the vertical position scan are presented in Fig.~\ref{fig:hor}(b). The optimal region for maximizing the FoM lies between 0~mm and 0.5~mm. Outside this range, the FoM decreases sharply, which is consistent with the reasons discussed previously. All simulation tools show good agreement with the measurement data, though \texttt{EGS5 + GPT} tends to underestimate the FoM.

\subsection{RF phase scan of accelerating structures in the capture linac}

The final measurement involved scanning the RF phases of the klystron units KL15 and KL16 in the capture linac. In a conventional accelerating linac, the RF phase is typically optimized to maximize energy gain. However, the RF phases preferred to be in a deceleration mode for $e^{+}$, particularly in the early acceleration stages. This configuration enhances $e^{+}$ bunching within a single RF bucket, reducing the energy spread of the $e^{+}$ bunch and thereby increasing the number of $e^{+}$ accepted by the DR.
The operational RF phases were determined relative to the zero-crossing phases, expressed as:
\begin{equation}
\phi_{\text{operational}} = \phi_{\text{zero-crossing}} + \phi_{\text{offset}}
\end{equation}
\noindent
where the zero-crossing RF phase corresponds to the setting at which the RF field induces neither acceleration nor deceleration of the beam, and $\phi_{\text{offset}}$ is the RF phase offset applied relative to the zero-crossing phase to achieve the desired operating mode (e.g., acceleration or deceleration). This method helps in setting a reference point for the comparison between the measurement and simulation. 

\begin{figure}[h!]
\centering
\includegraphics[width=\columnwidth]{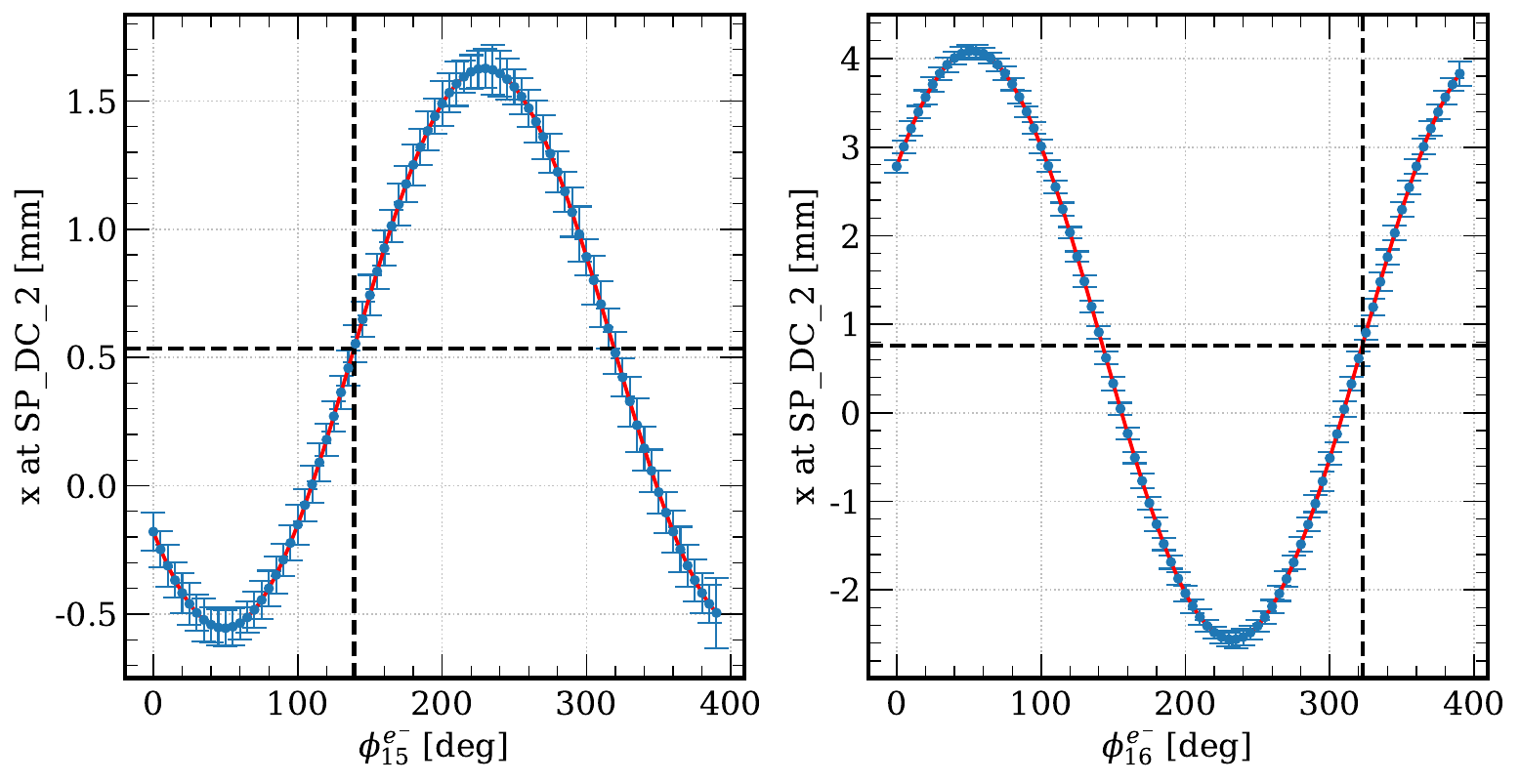}
\caption{\label{fig:z_cross} $e^{-}$ beam excursion measured in a high-dispersion region as a function of the RF phase for KL15 (left) and KL16 (right). The scan was performed in $5^\circ$ steps, with each measurement point averaged over 50 acquisitions to improve statistical accuracy. Error bars in the figure indicate the resulting standard deviations. The zero-crossing RF phases were determined by fitting the measured data with a cosine-like function. Dashed lines indicate the identified zero-crossing phases, given by $\phi_{15}^{0} = 139.1^\circ$ and $\phi_{16}^{0} = 322.5^\circ$.}
\end{figure}

The measurement began by determining the zero-crossing RF phases. These phases were identified by sending an $e^{-}$ beam through the $e^{-}$ hole in the target system (see Fig.~\ref{fig:target}) and scanning the RF phase of the first klystron unit KL15. The beam position was then recorded using an SP\_DC\_2 BPM located at the center of a large-dispersion region within the $e^{-}$ chicane upstream of Sector~3 (see Fig.~\ref{fig:KEK_injector}). Then, the measured positions were fitted with a cosine-like function to determine the zero-crossing phase. Once this phase was found for KL15, it was fixed, and the same procedure was repeated for KL16. The results of the zero-crossing phase measurements are presented in Fig.~\ref{fig:z_cross}. This method provided the RF phases for both klystron units $\phi_{15}^{0} = 139.1^\circ$
 and $\phi_{16}^{0} = 322.5^\circ$  where the net energy gain is zero. Note that the zero-crossing RF phases are the same for the $e^{-}$ and $e^{+}$ beams. These two RF phases serve as the reference points for subsequent RF phase scans. Since KL16 powers four LAS structures, it provides a higher energy gain than KL15. As a result, a larger $e^{-}$ beam excursion was observed, reaching a maximum of 4~mm for KL16, compared to 1.6~mm for KL15, as shown in Fig.~\ref{fig:z_cross}

\begin{figure}[H]
\centering
\includegraphics[width=0.91\columnwidth]{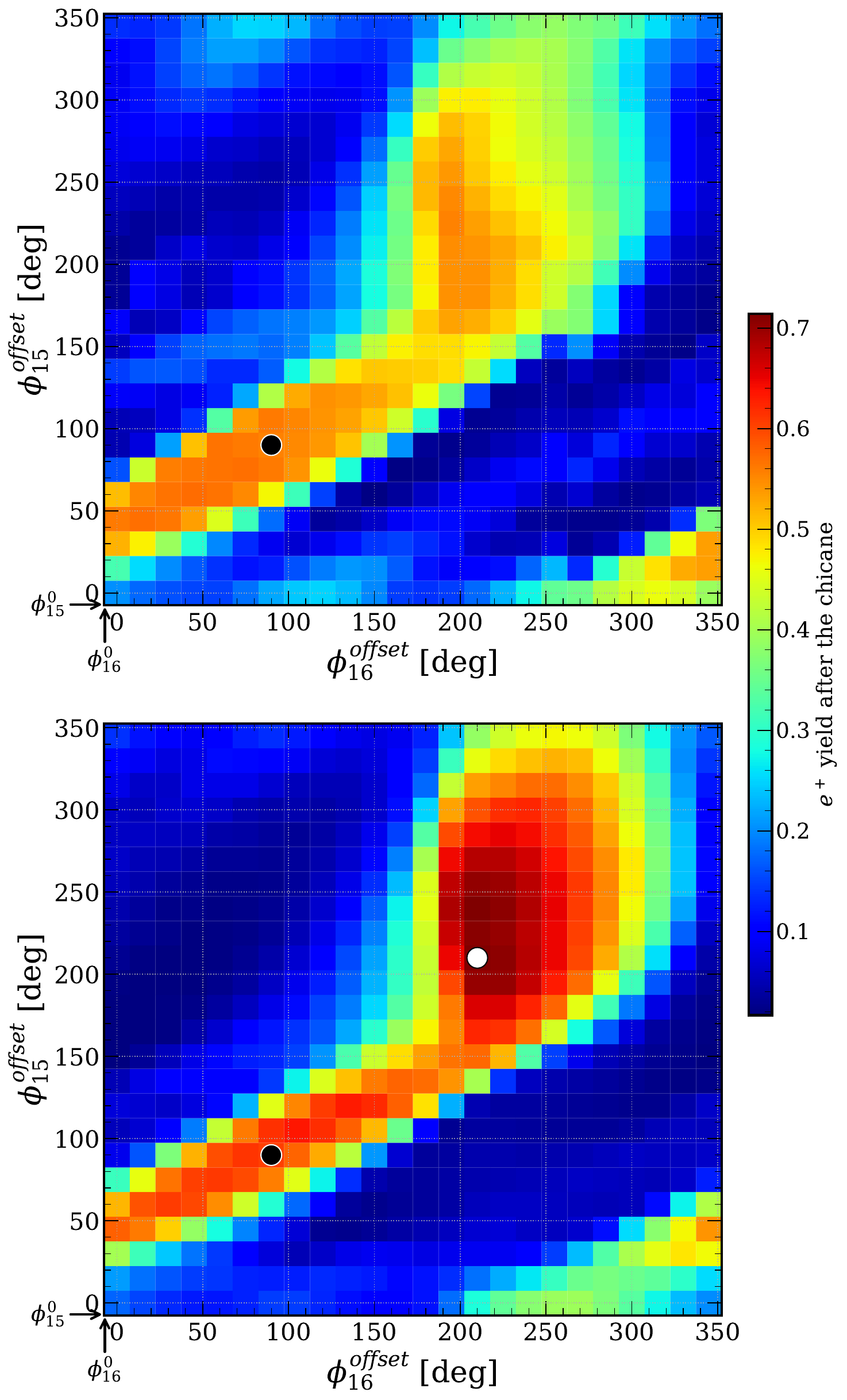}
\caption{\label{fig:2dphase_scan} Two-dimensional RF phase scan of the two klystron units KL15 and KL16 in the capture linac. Measurement data (top), in comparison with simulation results (bottom). Black markers indicate the working phases (deceleration mode)  $\Phi_{15,16}^{\text{operational}} = (229.1^\circ,52.5^\circ)$ at $\phi_{15,16}^{\text{offset}} = 90^\circ$. The white marker denotes the RF phase ($\phi_{15,16}^{\text{offset}} = 210^\circ$) corresponding to the simulated maximum $e^{+}$ yield of 0.72.}
\end{figure}

Starting from the zero-crossing RF phases, a two-dimensional RF phase scan was performed with a step of $10^\circ$ to map the RF phase response and identify the optimal operational point. The measurement data, shown in Fig.~\ref{fig:2dphase_scan}, are compared with simulation results obtained from our model.

\begin{figure}[h]
\centering
\includegraphics[width=0.91\columnwidth]{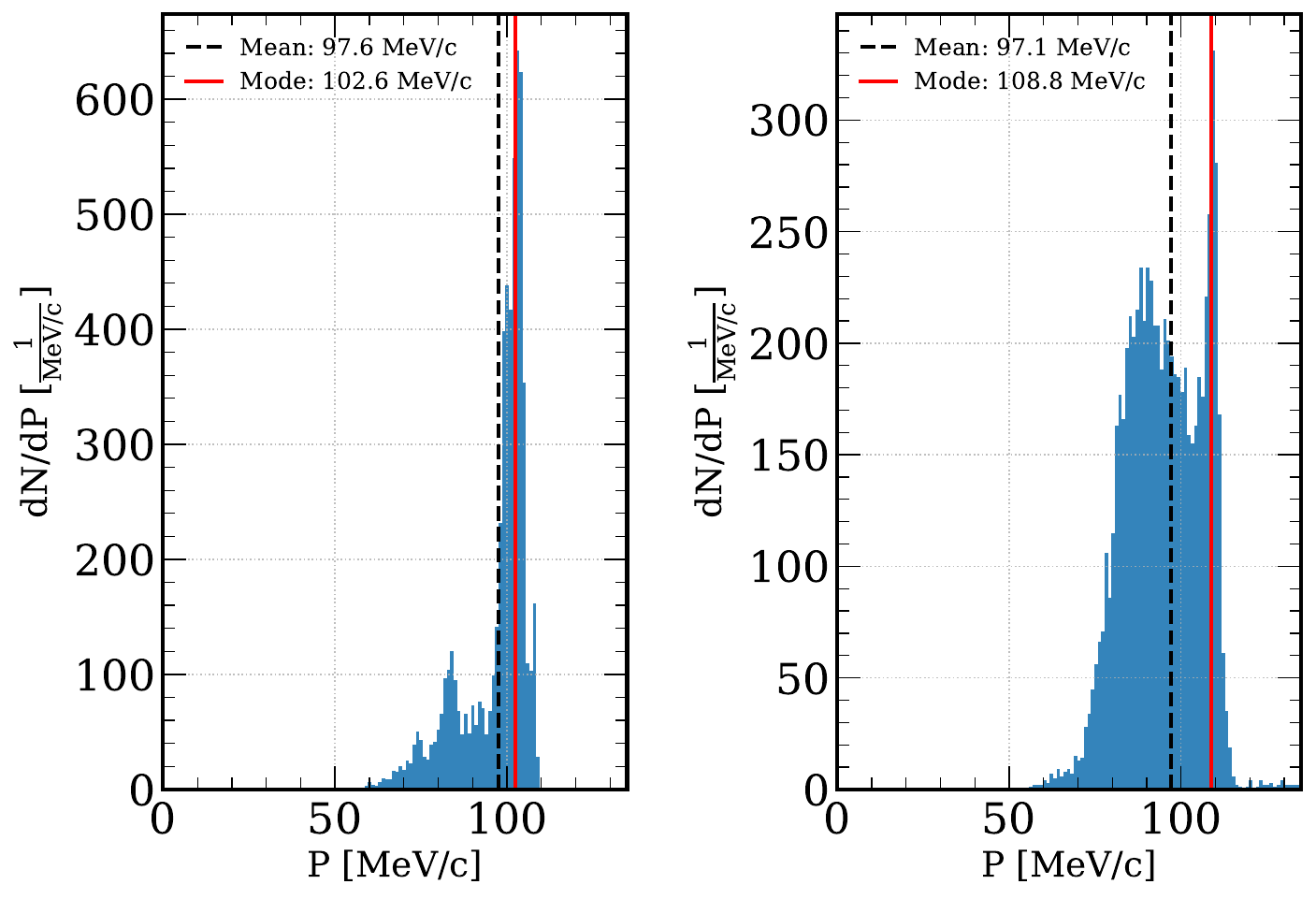}
\caption{\label{fig:E_spread} $e^{+}$ momentum distribution at the end of the capture section. The left plot corresponds to the operational RF phase (deceleration mode noted by the black marker in Fig.~\ref{fig:2dphase_scan}) $\phi_{15,16}^{\text{offset}} = 90^\circ$, while the right plot represents the RF phase that maximizes the $e^{+}$ yield at $\phi_{15,16}^{\text{offset}} = 210^\circ$ , (white marker on Fig.~\ref{fig:2dphase_scan}).}
\end{figure}

This scan identifies the deceleration RF phase at $\phi_{15,16}^{\text{offset}} = 90^\circ$ corresponding to operational phases of $\Phi_{15,16}^{\text{operational}} = (229.1^\circ,52.5^\circ)$, at this setting, the obtained $e^{+}$ yield ($N_{e^{+}}/N_{e^{-}}$) from measurement and simulation is 0.59 and 0.60, respectively.  
Meanwhile,  the acceleration RF phase at $\phi_{15, 16}^{\text{offset}} = 270^\circ$ corresponding to $\Phi_{15,16}^{\text{operational}} = (49.1^\circ,232.5^\circ)$ results in an $e^{+}$ yield of 0.46 from measurements and 0.6 from simulations. The black markers in Fig.~\ref{fig:2dphase_scan} indicate the selected working point corresponding to the deceleration mode. Although this RF phase does not provide the highest $e^{+}$ yield after the chicane, it significantly reduces the momentum spread and maximizes the accepted $e^{+}$ yield at the DR. At the chosen working RF phase, the momentum spread is 21.1~MeV/c, whereas, at the RF phase providing the highest (white marker on Fig.~\ref{fig:2dphase_scan} ) $e^{+}$ yield, it increases to 44.2~MeV/c, based on the simulation as shown in Fig.~\ref{fig:E_spread}. That is why the decelerated mode is adopted in the actual operation.

\begin{figure}[h!]
\centering
\includegraphics[width=\columnwidth]{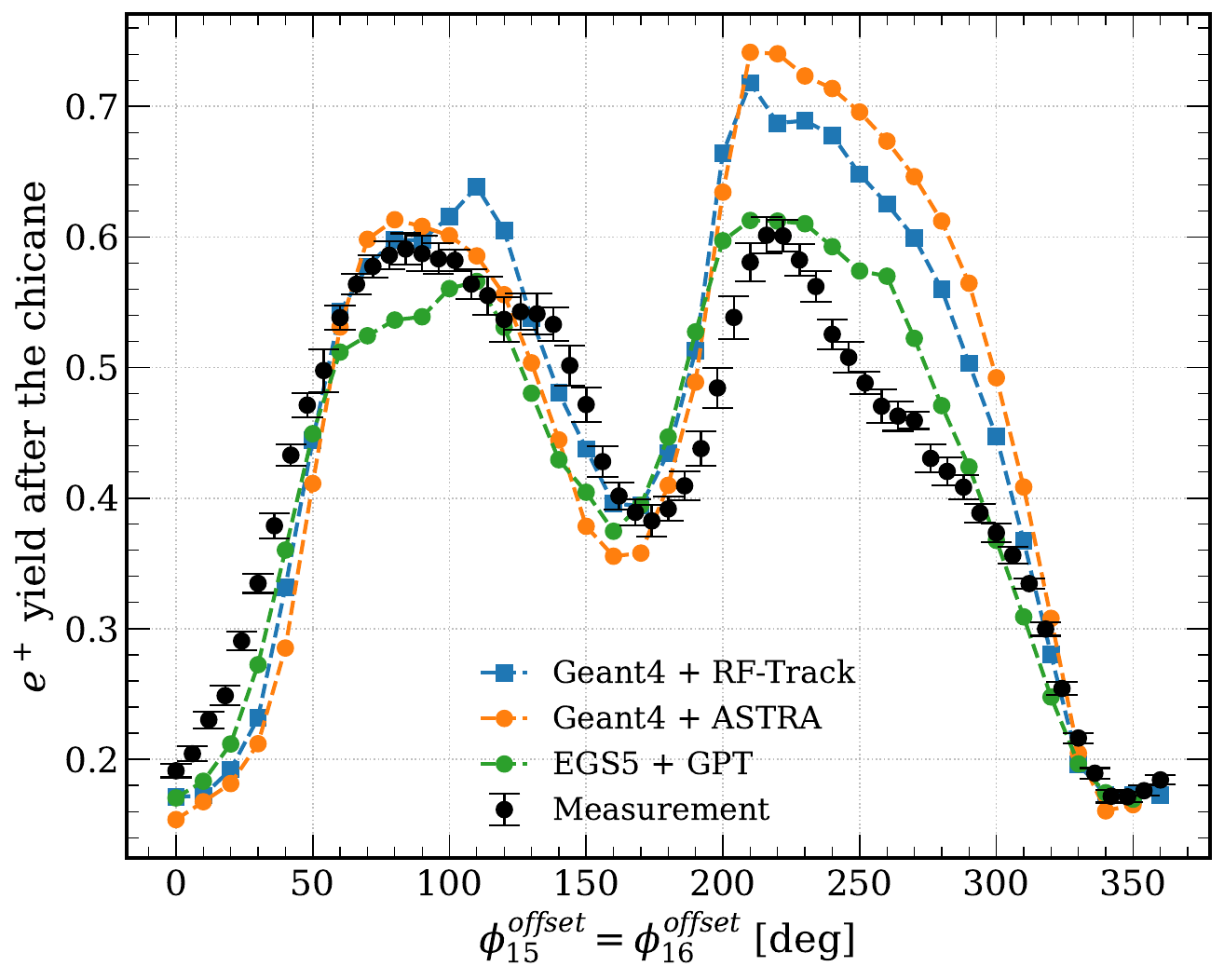}
\caption{\label{fig:phase_scan} $e^{+}$ yield as a function of the RF phase offset for KL15 and KL16.
The scan was performed in $5^\circ$ steps, with each measurement point averaged over 50 acquisitions to improve statistical accuracy. Error bars in the figure indicate the resulting standard deviations. The simulation data points are connected as a guide to the eye and do not represent a fit or model. }
\end{figure}

While this two-dimensional RF phase mapping provides valuable insight into the dependence of $e^{+}$ yield on the RF phases, it is not the most practical method for optimizing the capture section during operation and beam tuning studies. As observed in Fig.~\ref{fig:2dphase_scan}, the maximum $e^{+}$ yield is concentrated along the diagonal line, where the two RF phases are nearly equal. This suggests that a one-dimensional RF phase scan, rather than a full two-dimensional scan, is sufficient for practical optimization of the capture linac. 
The measurement was repeated using a one-dimensional RF phase scan with a finer step size of $5^\circ$ to refine the optimization process. The simulation results were then compared with the experimental data as illustrated in Fig.~\ref{fig:phase_scan}. 

In the first half of the scan, which corresponds to RF phases near the deceleration mode, there is a very good agreement between the simulation results and the measurements. However, in the second half of the scan, deviations between simulation and measured data become apparent. This discrepancy may be partly due to the idealized layout of the simulation, which does not account for alignment errors or magnetic field asymmetries. In particular, the DC solenoids in the capture linac have an extra dipole component arising from the winding of the hollow conductor, which is not included in the simulation. Further work is in progress to better understand these differences, and additional experimental and simulation studies are planned.

\section{conclusions}\label{sec:conclusion}

This paper presented an overview of the high-intensity $e^{+}$ sources, addressing the challenges associated with each stage and highlighting the complexity of $e^{+}$ source modeling. 
We have developed and experimentally validated a start-to-end simulation framework that couples \texttt{Geant4} for particle–matter interactions with \texttt{RF-Track} for 6D beam-dynamics tracking. Applied to SuperKEKB, the model reproduces the operational $e^{+}$ yield at the exit of the capture section. 
The start‑to‑end simulation predicts the $e^{+}$ yield of $0.6034~N_{e^{+}}/N_{e^{-}}$, only $0.0001$ (0.02~\%) higher than the measured value of $0.6033~N_{e^{+}}/N_{e^{-}}$, with statistical errors below 0.5~\% in both cases. The final experimental accuracy is set by the systematic uncertainty of the BPM charge calibration (typically a few percent), rather than by these statistical errors. 
Accordingly, the 0.02~\% simulation–measurement difference represents excellent agreement and is negligible compared with the overall error budget, demonstrating that the model accurately reproduces the $e^{+}$ source performance within the experimental uncertainties.
Furthermore, our model has been validated through several measurements conducted at the SuperKEKB $e^{+}$ source and compared with two independent models. The first model used at KEK is based on  \texttt{EG5 + GPT}, and the second model integrates \texttt{Geant4 + ASTRA}. The validation encompassed multiple key aspects of the $e^{+}$ source, including scanning of the primary $e^{-}$ impacting position on the target, the magnetic field of the solenoid in the capture section, and the RF phases in the capture linac. 
We analyzed the effects of these parameters on the $e^{+}$ yield at the end of the capture section, and the benchmarking study demonstrated good agreement with both measurements and the other simulation tools. 
This validated model establishes a solid foundation for application in other collider projects, providing a robust framework for designing and optimizing high-intensity $e^{+}$ sources.

\begin{acknowledgments}
We gratefully acknowledge the beamtime provided by the injector facility at KEK (Tsukuba, Japan). We thank Andrea Latina and Yongke Zhao for their valuable input on simulations in RF-Track.
Work is supported by ANR (Agence Nationale de la Recherche) under Grant No: ANR-21-CE31-0007 and the European Union’s Horizon 2020 Research and Innovation programme under Grant Agreement No 101004730.
The research leading to these results has received funding from the European Union's Horizon Europe MSCA under grant agreement no. 101086276

\end{acknowledgments}

\bibliographystyle{unsrt}
\bibliography{bibliography}

\begin{thebibliography}{10}

\bibitem{Aicheler2012}
M.~Aicheler et~al.
\newblock {\em A Multi-TeV Linear Collider Based on CLIC Technology: CLIC Conceptual Design Report}.
\newblock CERN Yellow Reports, Monographs. CERN, Geneva, 2012.

\bibitem{Benedikt2019}
M.~Benedikt et~al.
\newblock {FCC}-ee: the lepton collider.
\newblock {\em Eur. Phys. J. Spec. Top.}, 228:261--623, 2019.

\bibitem{Adolphsen:2013jya}
C.~Adolphsen et~al.
\newblock {The International Linear Collider Technical Design Report - Volume 3.I: Accelerator \& in the Technical Design Phase}.
\newblock 6 2013.

\bibitem{CEPCStudyGroup2023}
The CEPC~Study Group.
\newblock {CEPC} technical design report -- accelerator (v2).
\newblock December 2023.
\newblock \url{https://arxiv.org/abs/2312.14363}.

\bibitem{FS_FCC}
M.~Benedikt et~al.
\newblock Future circular collider feasibility study report volume 2: Accelerators, technical infrastructure and safety, 2025.

\bibitem{Alharthi:2025gpf}
F.~Alharthi et~al.
\newblock {{FCC}-ee positron source from conventional to crystal-based}.
\newblock 2 2025.

\bibitem{Chaikovska_2022}
I.~Chaikovska et~al.
\newblock Positron sources: from conventional to advanced accelerator concepts-based colliders.
\newblock {\em Journal of Instrumentation}, 17(05):P05015, may 2022.

\bibitem{NLC:2001aa}
{2001 Report on the Next Linear Collider: A Report submitted to Snowmass 2001}.
\newblock In {\em {APS / DPF / DPB Summer Study on the Future of Particle Physics}}, 6 2001.

\bibitem{Reuter:trolling}
E.~M. Reuter et~al.
\newblock {Mechanical design and development of a high power target system for the SLC positron source}.
\newblock {\em Conf. Proc. C}, 910506:1999--2001, 1991.

\bibitem{omori2024}
T.~Omori et~al.
\newblock Development of rotating target with ferrofluid seal for ilc electron-driven positron source, 2024.

\bibitem{Chehab1989}
R.~Chehab et~al.
\newblock Study of a positron source generated by photons from ultrarelativistic channeled particle.
\newblock In {\em Proceedings of the 1989 IEEE Particle Accelerator Conference (PAC'89)}, page 283, Chicago, IL, USA, 1989. IEEE.

\bibitem{Zhao:2025fzn}
Y.~Zhao, S.~Doebert, and A.~Latina.
\newblock {Performance optimization of the {CLIC} positron source}.
\newblock {\em Phys. Rev. Accel. Beams}, 28(1):011002, 2025.

\bibitem{chehab_positron_1989}
R.~Chehab.
\newblock {Positron sources}.
\newblock Technical Report LAL-RT-89-02, 1989.

\bibitem{agostinelli2003geant4}
S.~Agostinelli et~al.
\newblock Geant4—a simulation toolkit.
\newblock {\em Nuclear instruments and methods in physics research section A: Accelerators, Spectrometers, Detectors and Associated Equipment}, 506(3):250--303, 2003.

\bibitem{allison2006geant4}
J.~Allison et~al.
\newblock Geant4 developments and applications.
\newblock {\em IEEE Transactions on nuclear science}, 53(1):270--278, 2006.

\bibitem{allison2016recent}
J.~Allison et~al.
\newblock Recent developments in {G}eant4.
\newblock {\em Nuclear instruments and methods in physics research section A: Accelerators, Spectrometers, Detectors and Associated Equipment}, 835:186--225, 2016.

\bibitem{fluka1}
G.~Battistoni et~al.
\newblock Overview of the {FLUKA} code.
\newblock {\em Annals of Nuclear Energy}, 82:10--18, 2015.
\newblock Joint International Conference on Supercomputing in Nuclear Applications and Monte Carlo 2013, SNA + MC 2013. Pluri- and Trans-disciplinarity, Towards New Modeling and Numerical Simulation Paradigms.

\bibitem{fluka2}
C.~Ahdida et~al.
\newblock New capabilities of the {FLUKA} multi-purpose code, jan 2022.

\bibitem{EGS5}
H.~Hirayama et~al.
\newblock {SLAC-R-730}.
\newblock Technical Report SLAC-R-730, SLAC, 2005.
\newblock Also published as KEK Report 2005-8.

\bibitem{CSTStudioSuite}
{CST - Computer Simulation Technology}.
\newblock Cst studio suite, 2020.
\newblock Version 2020, a commercial electromagnetic simulation software.

\bibitem{ansys}
{ANSYS, Inc.}
\newblock Ansys.
\newblock \url{https://www.ansys.com/}.

\bibitem{astra}
Klaus Floettmann.
\newblock {ASTRA} code.
\newblock \url{https://www.desy.de/~mpyflo/}, 2017.

\bibitem{Elegent}
M.~Borland.
\newblock {ELEGANT: A flexible SDDS-compliant code for accelerator simulation}.
\newblock Technical report, Argonne National Lab., IL (US), 08 2000.

\bibitem{MadX}
L.~Deniau et~al.
\newblock Methodical{A}ccelerator{D}esign/{MAD-X}: 5.09.01, December 2023.

\bibitem{sad}
{KEK Accelerator Physics Group}.
\newblock {SAD — Strategic Accelerator Design}.

\bibitem{RFT}
Andrea Latina.
\newblock R\uppercase{F}-\uppercase{T}rack reference manual.
\newblock Technical report, CERN, Geneva, Switzerland, 2024.

\bibitem{SKEKB}
Y.~Ohnishi et~al.
\newblock Accelerator design at {S}uper{KEKB}.
\newblock {\em Progress of Theoretical and Experimental Physics}, 2013(3):03A011, 03 2013.

\bibitem{Ohnishi2023}
Y.~Ohnishi.
\newblock {SuperKEKB and Belle II}.
\newblock In {\em Proceedings of eeFACT2022}, pages 1--6. JACoW, 2023.

\bibitem{KEKNews2025}
{High Energy Accelerator Research Organization (KEK)}.
\newblock Super{KEKB} achieves new world record luminosity.
\newblock \url{https://www2.kek.jp/ipns/en/news/7015/}, 2025.

\bibitem{Natsui:2023jlc}
T.~Natsui et~al.
\newblock {KEK $e^+$/$e^-$ Injector Linac}.
\newblock {\em JACoW}, eeFACT2022:THYAT0102, 2023.

\bibitem{Kamitani2014}
T.~Kamitani et~al.
\newblock {SuperKEKB Positron Source Construction Status}.
\newblock In {\em Proceedings of IPAC’14}, pages 579--581, Dresden, Germany, Jun. 2014.

\bibitem{Zang2014}
L.~Zang et~al.
\newblock {Positron Yield Optimization by Adjusting the Components Offset and Orientation}.
\newblock In {\em Proceedings of IPAC’14}, pages 576--578, Dresden, Germany, Jun. 2014.

\bibitem{Enomoto2021}
Y.~Enomoto et~al.
\newblock {A New Flux Concentrator Made of Cu Alloy for the SuperKEKB Positron Source}.
\newblock In {\em Proceedings of IPAC’21}, pages 2954--2956, Campinas, Brazil, May 2021.

\bibitem{Matsumoto2014}
S.~Matsumoto et~al.
\newblock {Large-aperture Travelling-wave Accelerator Structure for Positron Capture of SuperKEKB Injector Linac}.
\newblock In {\em Proceedings of IPAC’14}, pages 3872--3874, Dresden, Germany, Jun. 2014.

\bibitem{Zhao_PEDD}
Y.~Zhao et~al.
\newblock Optimisation of the {CLIC} positron source at the 1.5 {TeV} and 3 {TeV} stages.
\newblock Technical report, CERN, Geneva, 2020.

\bibitem{superfish}
J.~H. Billen and L.~M. Young.
\newblock {POISSON/SUPERFISH} user's manual.
\newblock Technical Report LA-UR-96-1834, Los Alamos National Laboratory, 2006.
\newblock Los Alamos National Laboratory Report.

\end{thebibliography}

\end{document}